\documentclass[fleqn,usenatbib]{mnras}
\usepackage{newtxtext,newtxmath}
\usepackage[T1]{fontenc}

\DeclareRobustCommand{\VAN}[3]{#2}
\let\VANthebibliography\thebibliography
\def\thebibliography{\DeclareRobustCommand{\VAN}[3]{##3}\VANthebibliography}

\usepackage{graphicx}	% Including figure files
\usepackage{amsmath}	% Advanced maths commands
\usepackage{multicol}
\usepackage{subcaption}
\usepackage{upgreek}
\usepackage{tablefootnote}
\usepackage{threeparttable}
\title[Outflow of Q0254-334]{BAL Outflow in Quasar B0254-3327B: Analysis and Comparison with Other Extreme UV Outflows}

\author[D. Byun et al.]{
Doyee Byun,$^{1}$\thanks{E-mail:dbyun@vt.edu}
Nahum Arav,$^{1}$
Maryam Dehghanian,$^{1}$
Gwen Walker,$^{1}$
and Gerard A. Kriss$^{2}$
\\
$^{1}$Department of Physics, Virginia Tech, Blacksburg, VA 24061, USA\\
$^{2}$ Space Telescope Science Institute, 3700 San Martin Drive, Baltimore, MD 21218, USA
}

\date{Accepted XXX. Received YYY; in original form ZZZ}

\pubyear{2023}

\begin{document}
\label{firstpage}
\pagerange{\pageref{firstpage}--\pageref{lastpage}}
\maketitle

\begin{abstract}
We have identified a broad absorption line (BAL) outflow in the HST/STIS spectrum of the quasar QSO B0254-3327B at velocity $v=-3200\text{ km s$^{-1}$}$. The outflow has absorption troughs from ions such as \ion{Ne}{viii}, \ion{Na}{ix}, \ion{Si}{xii}, and \ion{Ne}{v}. We also report the first detection of \ion{S}{xiv} absorption troughs, implying very high ionization. Via measurement of the ionic column densities, photoionization analysis, and determination of the electron number density of the outflow, we found the kinetic luminosity of the outflow system to be up to $\sim1\%$ of the quasar's Eddington luminosity, or $\sim5\%$ of the bolometric luminosity, making it a potential contributor to AGN feedback. A solution with two ionization phases was needed, as a single phase was not sufficient to satisfy the constraints from the measured ionic column densities. We find that the ionization parameter of the very high-ionization phase of the outflow is within the expected range of an X-ray warm absorber. We also examined the physical properties of the outflow of Q0254-334 along with previously studied extreme UV outflows, with a total sample of 24 outflow systems, finding a weak negative correlation between outflow velocity and distance from the central source, with larger distances corresponding to slower velocities. The very high-ionization phase of the Q0254-334 outflow has one of the highest ionization parameters of UV absorption outflows to date, which we attribute to the presence of \ion{S}{xiv}.
\end{abstract}

\begin{keywords}
galaxies: active -- quasars: absorption lines -- quasars: individual: QSO B0254-3327B
\end{keywords}

%%%%%%%%%%%%%%%%%%%%%%%%%%%%%%%%%%%%%%%%%%%%%%%%%%

%%%%%%%%%%%%%%%%% BODY OF PAPER %%%%%%%%%%%%%%%%%%

\section{Introduction}

Quasar absorption outflows are often invoked as likely contributors to active galactic nucleus (AGN) feedback \citep[e.g.][]{1998A&A...331L...1S,2004ApJ...608...62S,2018ApJ...857..121Y,2021ApJ...919..122V,2022SciA....8.3291H}. They are detected via blueshifted absorption troughs in the rest frame of $\lesssim 40\%$ of quasars \citep[e.g.,][]{2003AJ....125.1784H,2006ApJS..165....1T,2008ApJ...672..108D,2008MNRAS.386.1426K,2011MNRAS.410..860A}. In order to contribute to AGN feedback, outflow systems theoretically require a kinetic luminosity ($\dot{E}_k$) of at least $\sim0.5\%$ \citep{2010MNRAS.401....7H} or $\sim5\%$ \citep{2004ApJ...608...62S} of the quasar's luminosity, which we interpret to be the Eddington luminosity ($L_{Edd}$) following the reasoning described by \citet{2020MNRAS.499.1522M}, as opposed to the bolometric luminosity ($L_{Bol}$). Past studies have found outflows that fit one or both of these criteria \citep[e.g.][]{2009ApJ...706..525M,2013ApJ...762...49B,2015MNRAS.450.1085C,2018ApJ...866....7L,2020ApJS..247...39M,2020ApJ...891...53C,2022ApJ...937...74C,10.1093/mnras/stac2638,2022Byun,2022MNRAS.516.3778W}.\par
In order to find the value of $\dot{E}_k$, it is important to find the mass flow rate ($\dot{M}$), a method for which involves finding the electron number density ($n_e$) and ionization parameter ($U_H$) to measure the distance ($R$) of the outflow from the central source \citep{2012ApJ...758...69B}. Multiple quasar outflows have been analyzed via this method \citep[e.g.][]{2001ApJ...548..609D,2001ApJ...550..142H,2022MNRAS.516.3778W,10.1093/mnras/stac2194}. For ionized outflows, the ionization parameter can be determined by measuring the column densities of ions, and comparing them with simulated values based on a range of $U_H$ and hydrogen column density ($N_H$). Multiple outflow analysis studies have been conducted using the spectral synthesis code \textsc{Cloudy} \citep{2017RMxAA..53..385F} for this method. \citep[e.g.][]{2018ApJ...858...39X,2020ApJS..247...39M,10.1093/mnras/stac2194,2022MNRAS.516.3778W}.\par
This paper presents the analysis of the absorption outflow of the quasar QSO B0254-3327B (hereafter Q0254-334), using the method described above, based on HST/STIS observational data, ultimately finding the ratio between $\dot{E}_k$ and $L_{Edd}$.\par
The paper is structured as follows. Section \ref{sec:observation} describes the observation and data acquisition of Q0254-334; section \ref{sec:analysis} discusses the process of finding the ionic column densities, $N_H$, $U_H$, and $n_e$ of the outflow; section \ref{sec:discussion} shows our analysis results of whether the outflow's kinetic luminosity is sufficient to contribute to AGN feedback, and compares our results with studies of other outflows; and section \ref{sec:conclusion} concludes and summarizes the paper. In our analysis, we adopted a cosmology of $h=0.696, \Omega_m=0.286$, and $\Omega_\Lambda=0.714$ \citep{Bennett_2014}. We used the \textsc{Python} astronomy package \textsc{Astropy} \citep{astropy:2013,astropy:2018} for our cosmological calculations. We also used \textsc{Scipy} \citep{2020SciPy-NMeth}, \textsc{Numpy} \citep{harris2020array}, and \textsc{Pandas} \citep[v1.2.4,][]{jeff_reback_2021_4681666,mckinney-proc-scipy-2010} for the majority of our numerical computations, as well as \textsc{Matplotlib} \citep{Hunter:2007} for plotting our figures. \par
\begin{figure*}
    \centering
    \includegraphics[width=\linewidth]{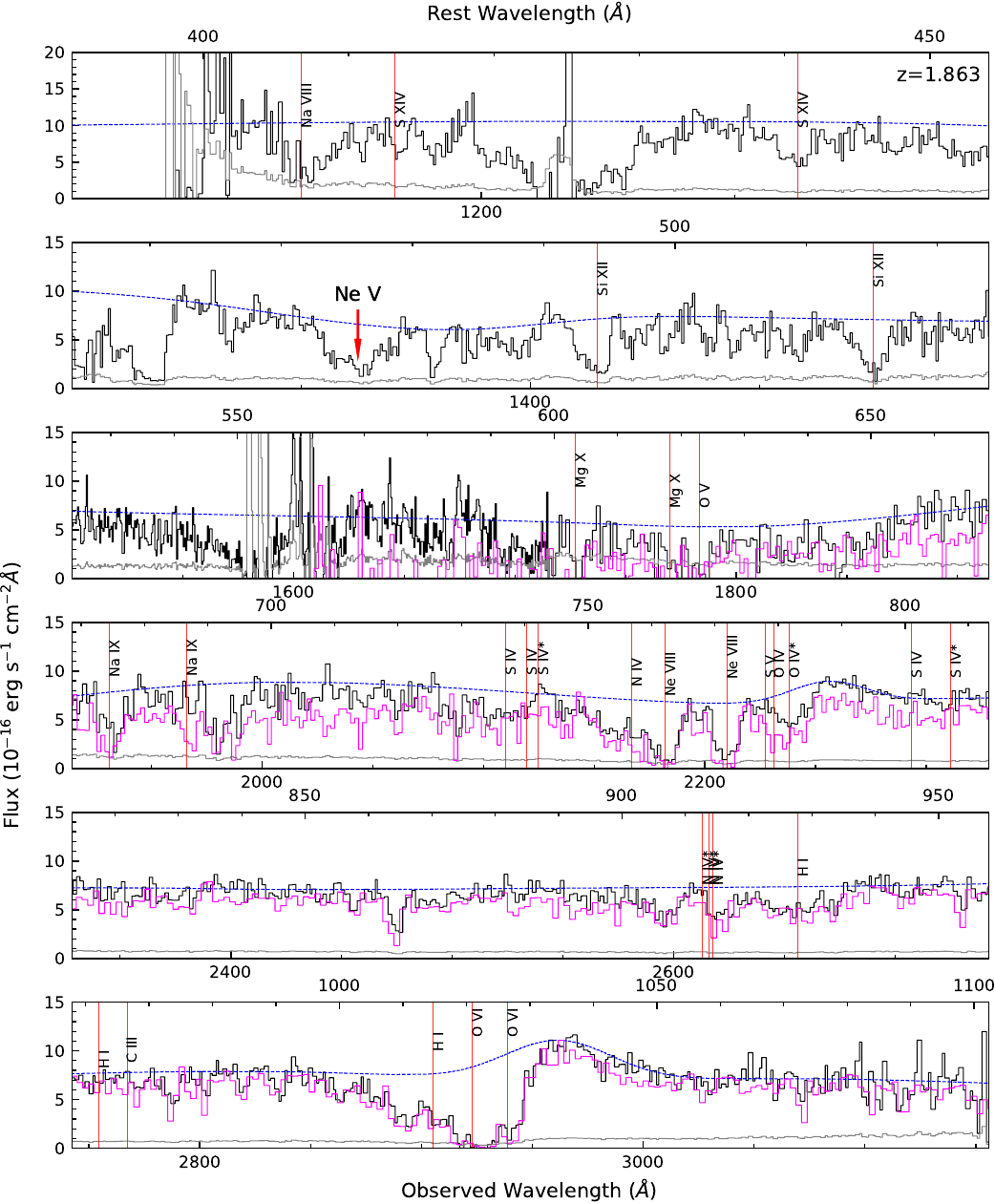}
    \caption{Co-Added STIS spectrum (black) and FOS spectrum (purple) of Q0254-334. The absorption features of the outflow system ($v\approx-3200\text{ km s}^{-1}$) are marked with red vertical lines, with the \ion{Ne}{v} multiplet emphasized with a red arrow. The continuum and emission model is plotted as a blue dashed curve. Note that the continuum flux has risen between 1993 and 2001 in observed wavelengths up to $\sim2100$\AA, while at longer wavlengths, the variability is nearly indistinguishable.}
    \label{fig:fluxplot}
\end{figure*}
\section{Observation and Data Acquisition}
\label{sec:observation}
Q0254-334 (J2000; RA=02:56:47.84, DEC=-33:15:26.16, z=1.863) was observed with HST/STIS on 17 February, 2001 as part of the program SNAP 8681, and on 4 March, 2001 as part of the program GO 8569, with the G230L and G140L gratings respectively. Prior to this, it was also observed with HST/FOS in 1994. Due to the limited wavelength range of the FOS data relative to that of STIS, we have focused on the STIS data for the purpose of this analysis. After retrieving the data from the Mikulski Archive for Space Telescopes, we have co-added the two STIS spectra, and corrected the combined spectrum for galactic reddening and extinction with $E(B-V)=0.0205$ \citep{2011ApJ...737..103S}, and the extinction model by \citet{1999PASP..111...63F}. The co-added and dereddened spectrum of the two observations, covering observed wavelengths 1138.6--3156.6\AA, along with the FOS spectrum, is shown in Figure~\ref{fig:fluxplot}.\par
\begin{figure}
    \centering
    \includegraphics[width=\linewidth]{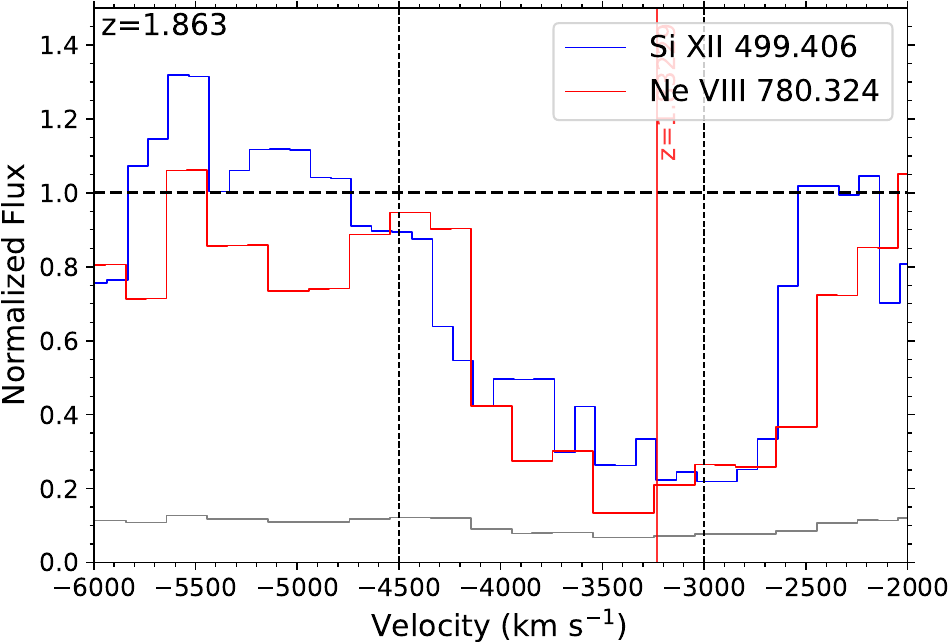}
    \caption{Normalized flux in the region of the \ion{Si}{xii} $\lambda499.406$ and \ion{Ne}{viii} $\lambda780.324$ absorption. The dashed horizontal line shows the continuum level, and the dotted vertical lines show the region between $v=-4500$ and $-3000$ km s$^{-1}$, which represents the range for the EUV BAL criteria, as defined by \citet{2020ApJS..247...37A}. The normalized flux for \ion{Si}{xii} falls below 0.9 in this region. While the \ion{Ne}{viii} normalized flux lies above 0.9 at $-4500$ km s$^{-1}$, it is within uncertainty range to fall below the threshold. To demonstrate, the error in the \ion{Ne}{viii} flux is shown in gray.}
    \label{fig:BAL_check}
\end{figure}

We have identified a broad absorption line (BAL) outflow system at $v=-3200\text{ km s}^{-1}$, with its ionic absorption troughs marked by red vertical lines in Figure~\ref{fig:fluxplot}. Troughs exist of species such as \ion{Ne}{viii}, \ion{Na}{ix}, and \ion{Si}{XII}, as well as excited state transitions such as \ion{O}{iv}* and \ion{Ne}{v}*. \citet{2020ApJS..247...37A} define a BAL in the extreme UV range as a continuous absorption feature with normalized flux $I\leq0.9$ over a width of $\Delta v \gtrsim 1500\text{ km s}^{-1}$, at least $-3000\text{ km s}^{-1}$ blueward of the center of emission. We have verified that the outflow is a BAL outflow by confirming the width of the \ion{Si}{xii} $\lambda499.406$ and \ion{Ne}{viii} $\lambda780.324$ troughs (see Figure~\ref{fig:BAL_check}). The normalized flux was found by modeling the quasar's continuum via a spline model that gave the minimum possible continuum above the absorption, and the most prominent emission features (e.g. \ion{O}{vi}) with Gaussians. The presence of the \ion{Ne}{v}* feature allowed us to find the value of $n_e$, as shown in Section~\ref{subsec:nev}.\par

\section{Analysis}
\label{sec:analysis}
\subsection{Ionic Column Densities}
As the ionic column densities $(N_{ion})$ of the outflow are crucial in finding the physical properties of the outflow, we used two different methods to find them based on the absorption troughs: the apparent optical depth (AOD) method in which we assume uniform and homogeneous covering \citep{Savage1991}; and the partial covering (PC) method in which we include a covering factor $C<1$ \citep{1997ASPC..128...13B,1999ApJ...524..566A,Arav1999}.\par
The AOD method allows us to find upper limits and lower limits of ionic column densities with its relative simplicity, while the PC method lets us find more accurate measurements of ions with doublet features \citep[e.g.][]{DeKool2002,2005ApJ...620..665A,2012ApJ...751..107B,10.1093/mnras/stac2638}. As done by \citet{2022Byun} for the quasar J024221.87+004912.6, we selected the appropriate method for computing the column density of each ion.\par
The AOD method involves the following relation between intensity and optical depth \citep{1978ppim.book.....S,Savage1991}:
\begin{equation}
    I(\lambda)=I_0 (\lambda)e^{-\tau(\lambda)}
\end{equation}
where $I(\lambda)$ is the intensity as a function of wavelength, $I_0(\lambda)$ is the intensity without absorption, and $\tau$ is the optical depth. Finding the optical depth enables computation of the column density, as they have the following relation:
\begin{equation}
    \tau(v)=\frac{\mathrm{\pi}e^2}{m_e c}f\lambda N(v)
\end{equation}
where $\tau(v)$ is the optical depth as a function of velocity, $e$ is the elementary charge, $m_e$ is the mass of an electron, and $N(v)$ is the column density per unit velocity. Integrating $N(v)$ over the velocity range of an ion's absorption trough results in the ion's column density.\par
As mentioned above, the PC method involves a covering factor $C<1$, which follows the relation shown in the equations below \citep{2005ApJ...620..665A}:
\begin{align}
    I_R(v)-[1-C(v)]=C(v)e^{-\tau(v)}\\
    I_B(v)-[1-C(v)]=C(v)e^{-2\tau(v)}
\end{align}
where $I_R(v)$ and $I_B(v)$ are the intensities at the red (longer wavelength) and blue (shorter wavelength) troughs of a doublet transition, $C(v)$ is the covering factor as a function of velocity, and $\tau$ is the optical depth.\par
\begin{figure*}
    \centering
    \begin{multicols}{3}
    \subcaptionbox{\ion{H}{i}\label{fig:HI}}{\includegraphics[width=0.33\textwidth]{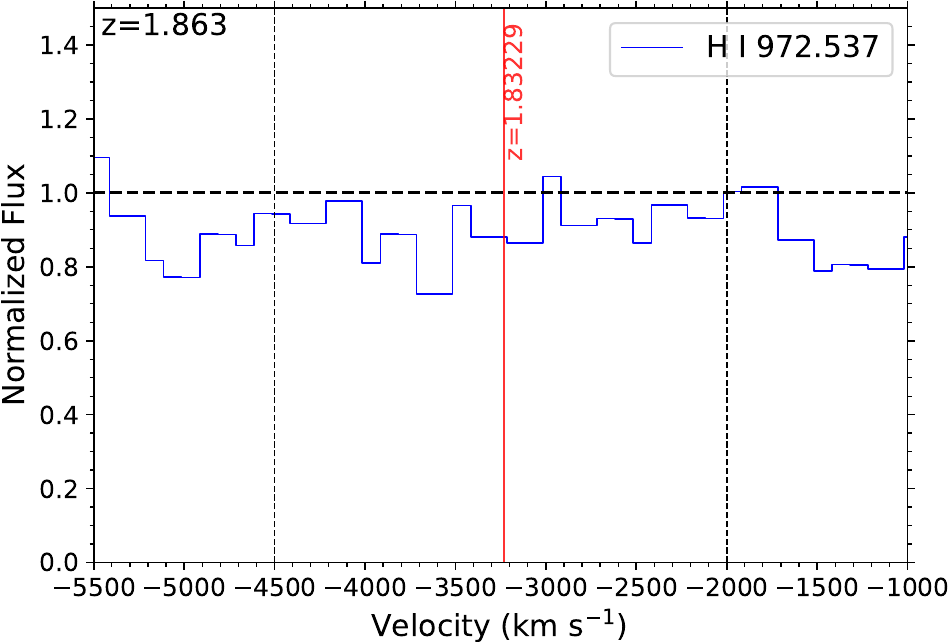}}\par
    \subcaptionbox{\ion{C}{iii}\label{fig:CIII}}{\includegraphics[width=0.33\textwidth]{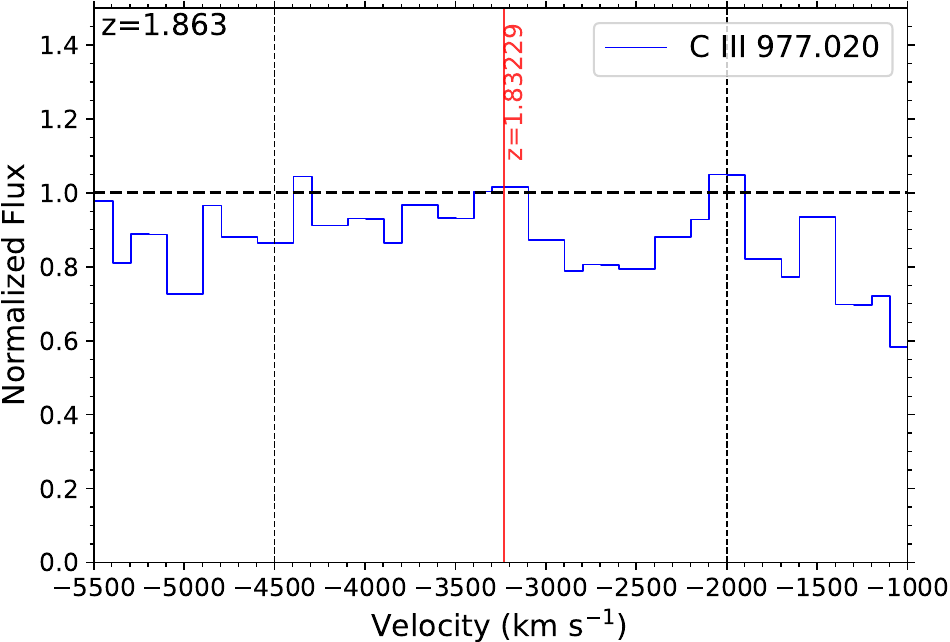}}\par
    \subcaptionbox{\ion{N}{iv}\label{fig:NIV}}{\includegraphics[width=0.33\textwidth]{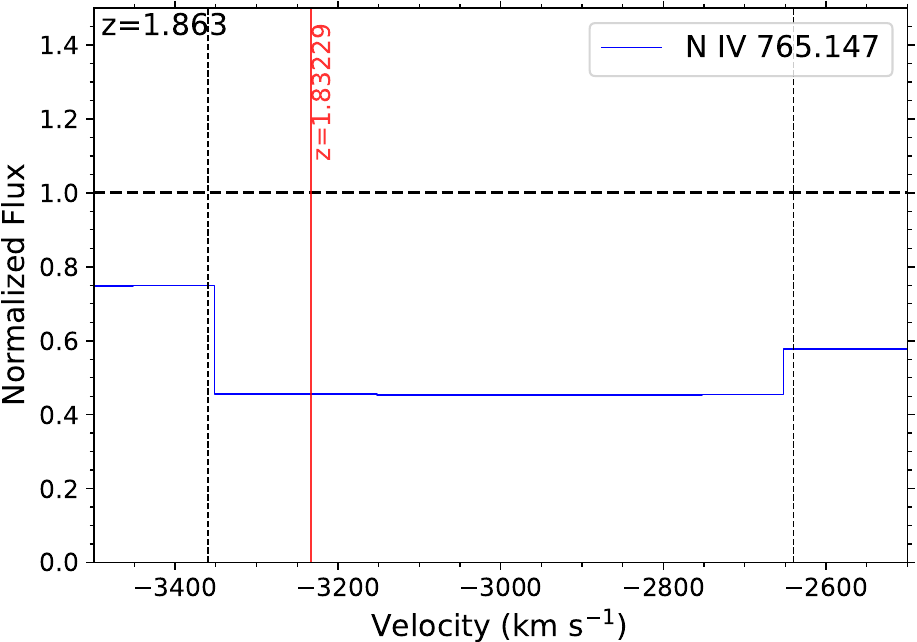}}\par
    \end{multicols}
    \begin{multicols}{3}
    \subcaptionbox{\ion{O}{iv}*\label{fig:OIV}}{\includegraphics[width=0.33\textwidth]{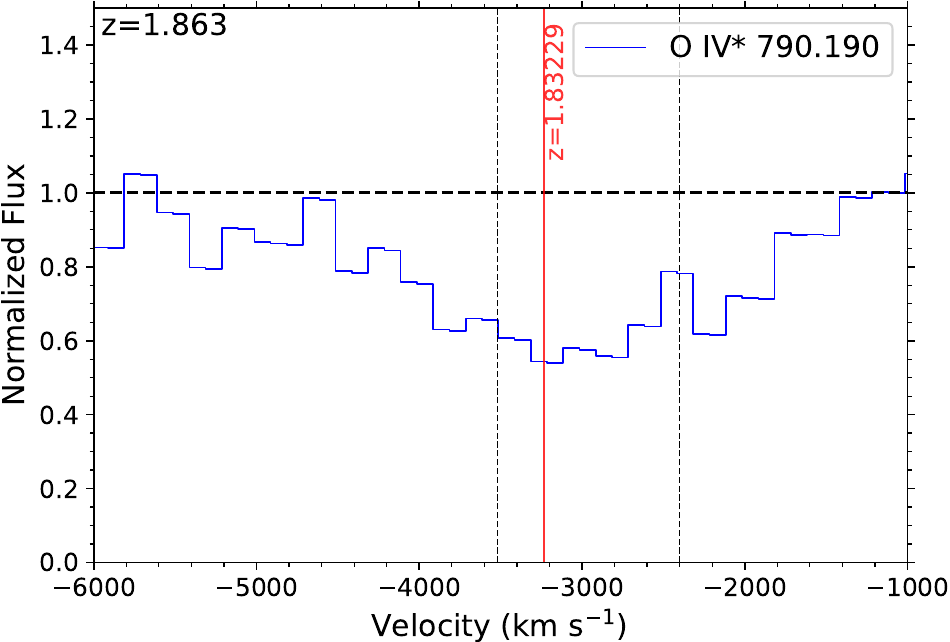}}\par
    \subcaptionbox{\ion{O}{v}\label{fig:OV}}{\includegraphics[width=0.33\textwidth]{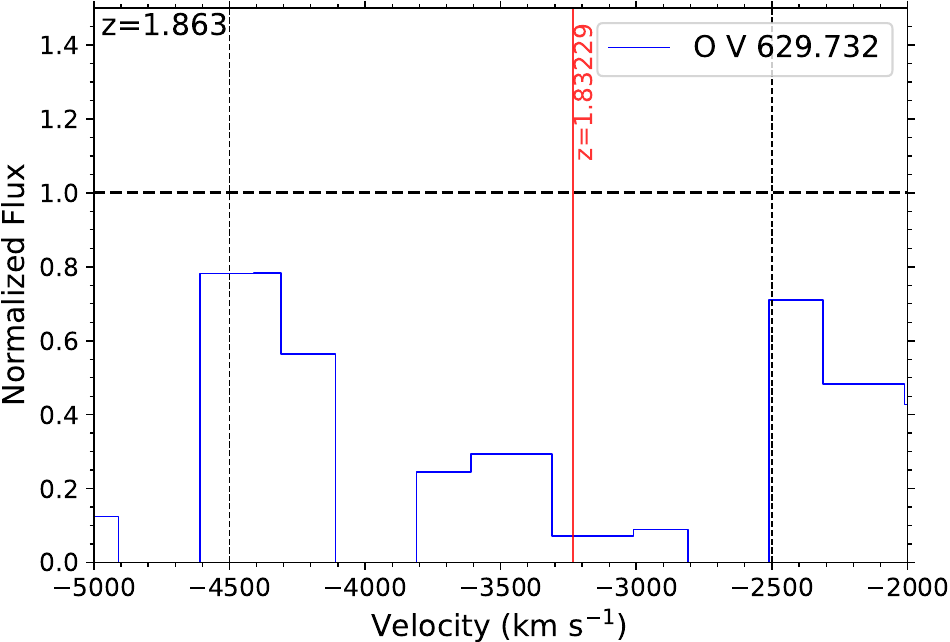}}\par
    \subcaptionbox{\ion{O}{vi}\label{fig:OVI}}{\includegraphics[width=0.33\textwidth]{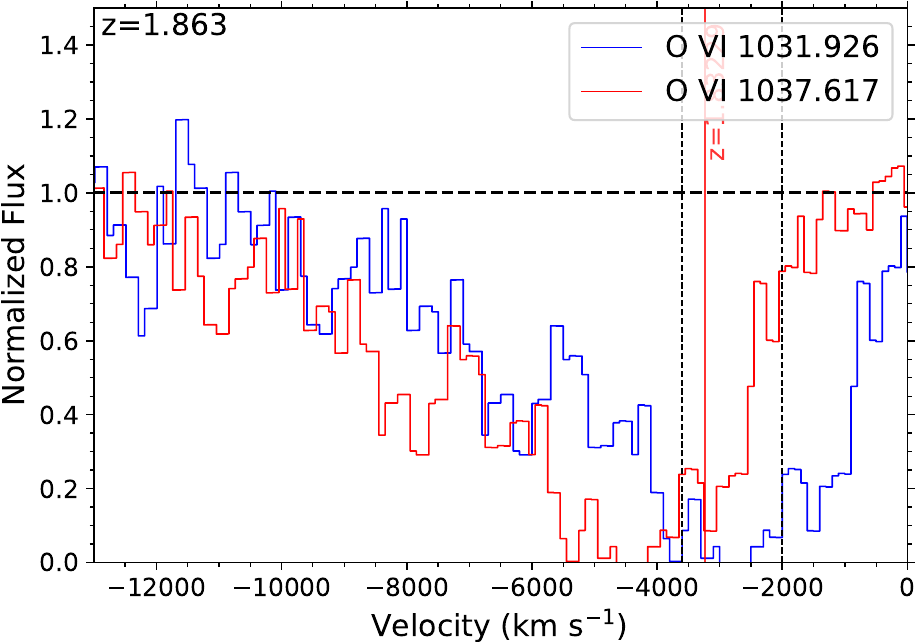}}\par
    \end{multicols}
    \begin{multicols}{3}
    \subcaptionbox{\ion{Ne}{viii}\label{fig:NeVIII}}{\includegraphics[width=0.33\textwidth]{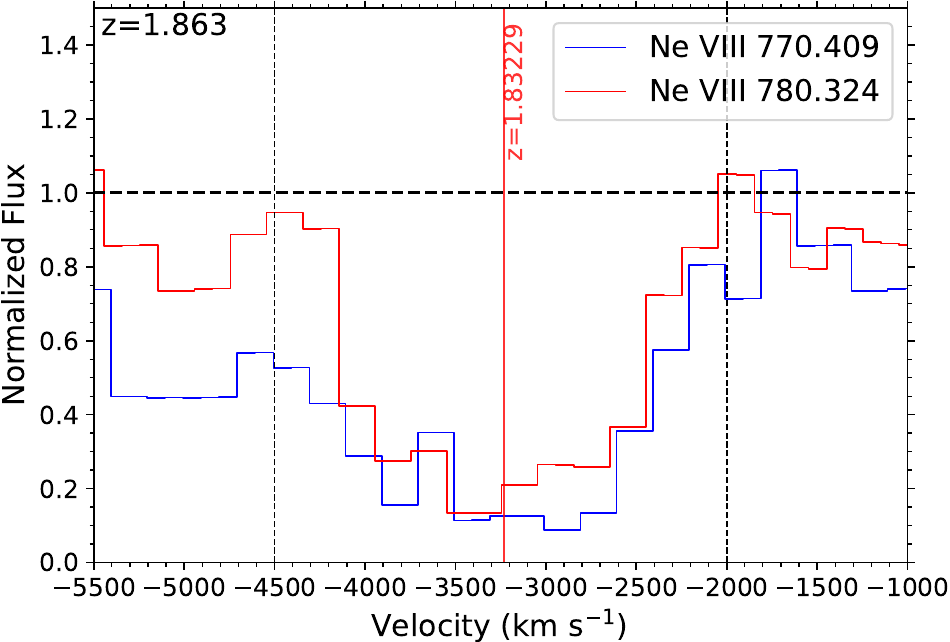}}\par
    \subcaptionbox{\ion{Na}{viii}\label{fig:NaVIII}}{\includegraphics[width=0.33\textwidth]{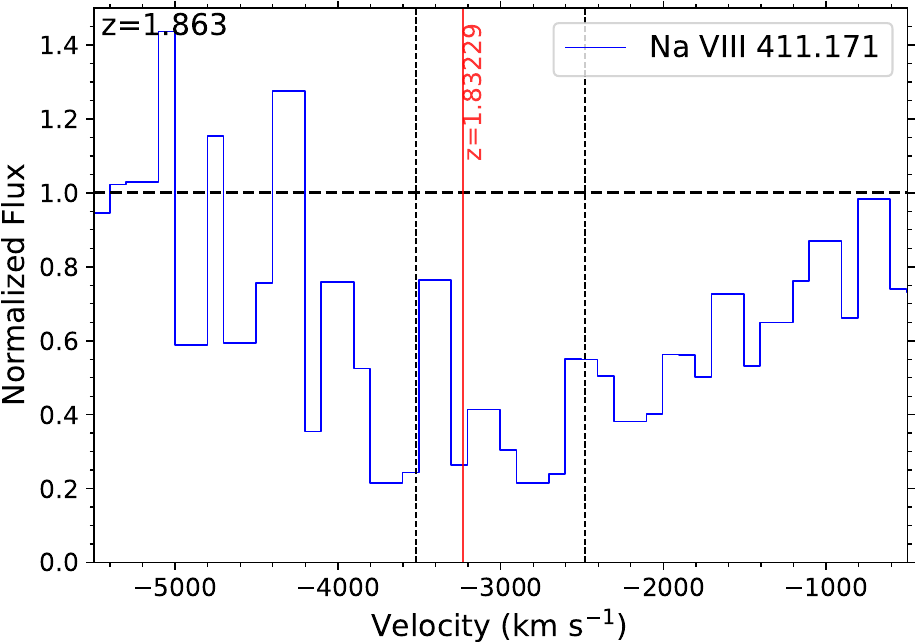}}\par
    \subcaptionbox{\ion{Na}{ix}\label{fig:NaIX}}{\includegraphics[width=0.33\textwidth]{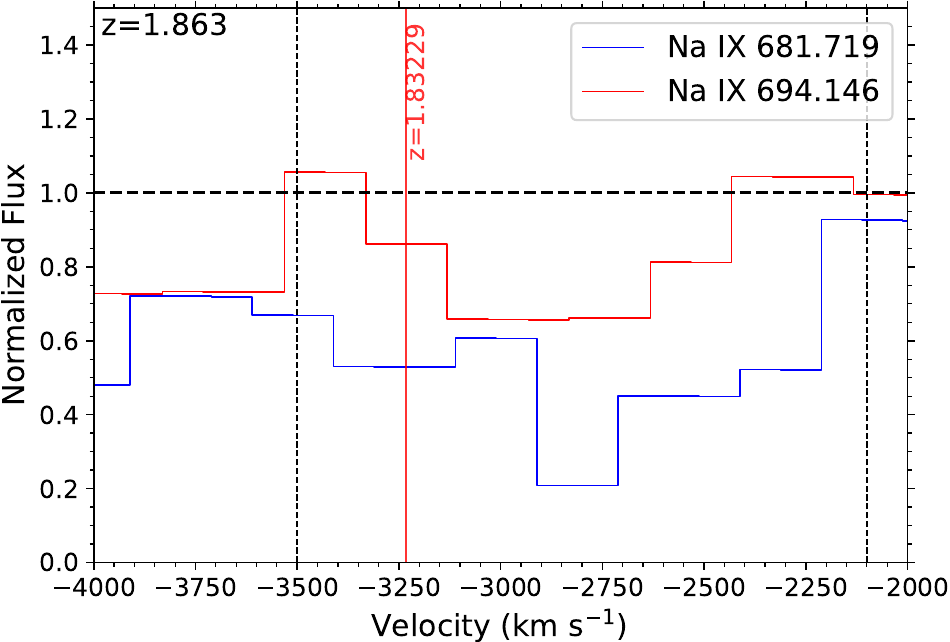}}\par
    \end{multicols}
\caption{Absorption troughs of the Q0254-334 outflow plotted in velocity space. The velocity of the outflow at $z=1.89229$ is marked with red vertical lines. The integration range used to calculate the column densities is marked with dotted vertical lines, while the continuum level is indicated by the dashed horizontal line.}
\label{fig:vcut}
\end{figure*}
\begin{figure*}
    \centering
    \begin{multicols}{2}
    \subcaptionbox{\ion{Mg}{x}\label{fig:MgX}}{\includegraphics[width=0.4\textwidth]{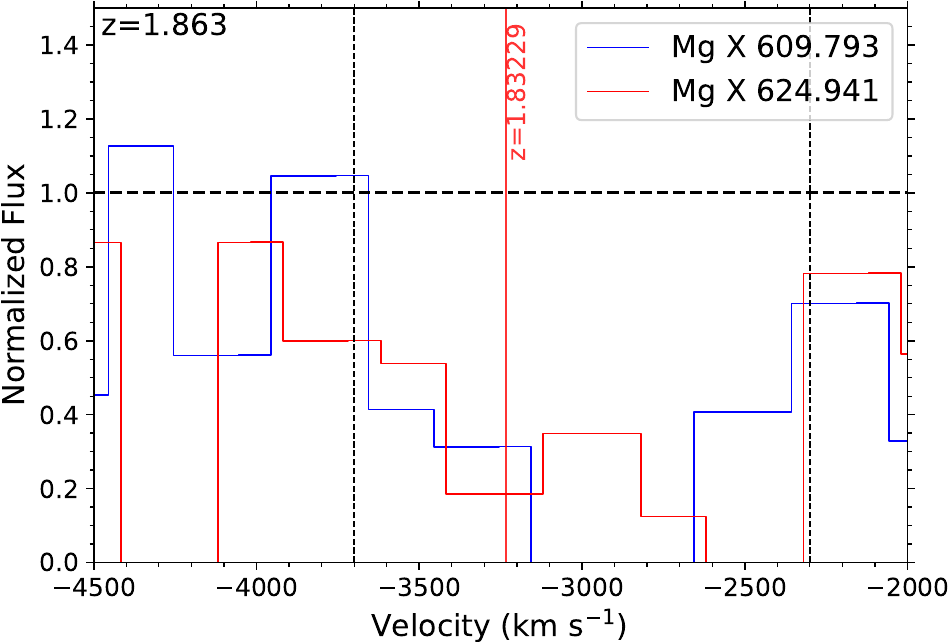}}\par
    \subcaptionbox{\ion{Si}{xii}\label{fig:SiXII}}{\includegraphics[width=0.4\textwidth]{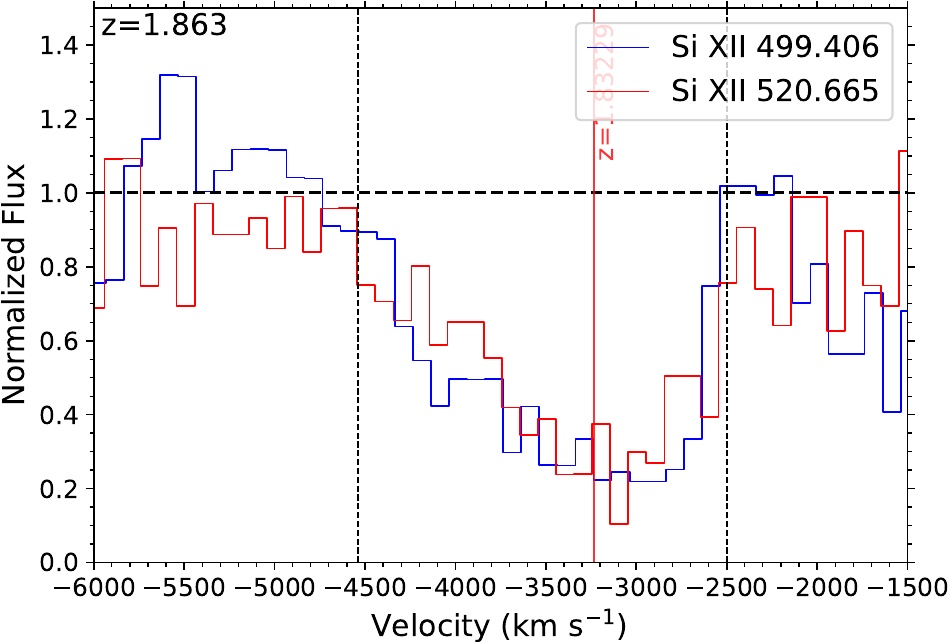}}\par
    \end{multicols}
    \begin{multicols}{2}
    \subcaptionbox{\ion{S}{xiv}\label{fig:SXIV}}{\includegraphics[width=0.4\textwidth]{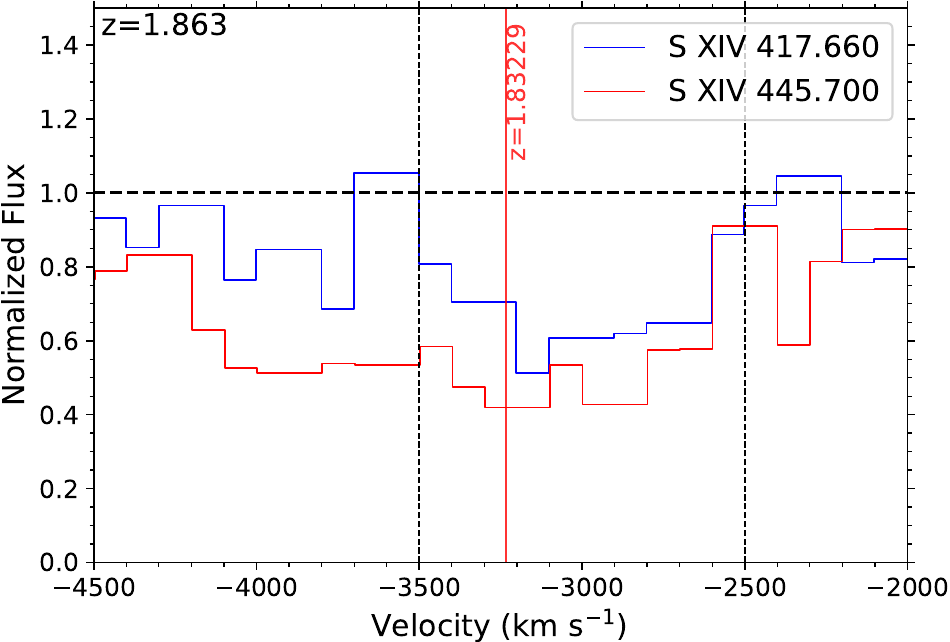}}\par
    \subcaptionbox{\ion{S}{iv}\label{fig:SIV*}}{\includegraphics[width=0.4\textwidth]{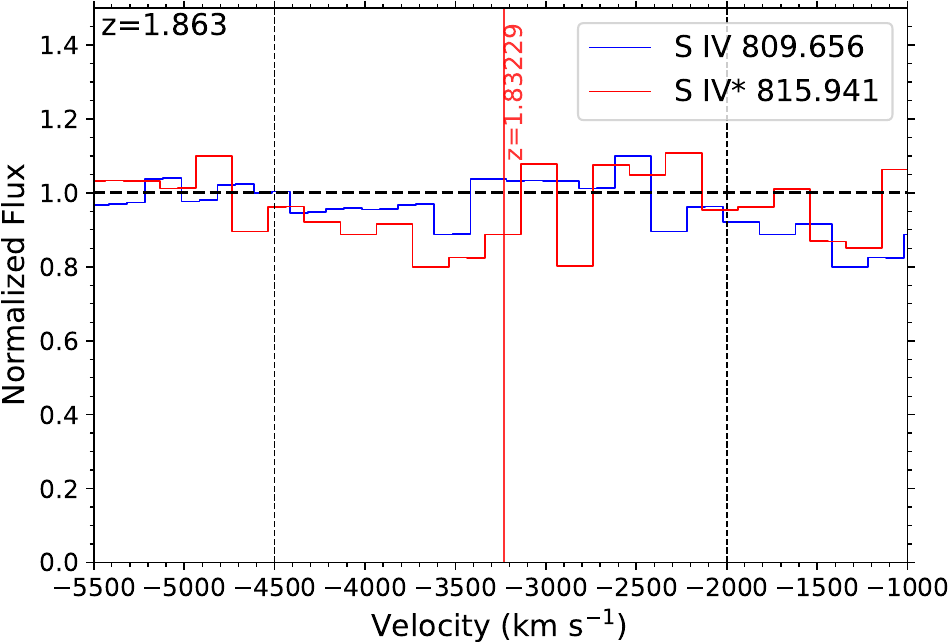}}\par
    \end{multicols}
\caption{Absorption troughs of \ion{Mg}{x}, \ion{Si}{xii}, \ion{S}{xiv}, and \ion{S}{iv}* in the outflow of Q0254-334. Format and notation are identical to those of Figure~\ref{fig:vcut}.}
\label{fig:vcut2}
\end{figure*}

For each ion, we converted the spectrum from wavelength space to velocity space, using the redshift of the quasar and the wavelengths of the ionic transition lines (see Figures~\ref{fig:vcut},\ref{fig:vcut2}). We then chose integration ranges for each ion that covered visible absorption features and minimized blending effects with other lines. For instance, the \ion{O}{vi} doublet had heavy blending between the red and blue troughs (see Figure~\ref{fig:OVI}). We thus chose a range where the overlap between the red and blue troughs would be minimized and computed a lower limit to the column density with the AOD method. As there were no discernible absorption troughs of \ion{Ly}{$\gamma$}, \ion{C}{iii}, and \ion{S}{iv*}, we measured their AOD column density with integration range $v\approx-4500$ to $-2000$ km s$^{-1}$ to match the \ion{Ne}{viii} width, and treated them as upper limits. Due to the severe blending in the multiplet of \ion{S}{iv} $\lambda\lambda 744.904, 748.393$ and \ion{S}{iv*} $\lambda750.221$ (see panel 4 of Figure~\ref{fig:fluxplot}), we were unable to pinpoint the column density of the resonance state \ion{S}{iv} 0 from this trough. However, as there was no discernible absorption trough of \ion{S}{iv} $\lambda809.656$, we were able to find an upper limit of its column density. Similarly, the trough of \ion{O}{iv} $\lambda787.711$ blended with \ion{O}{iv*} $\lambda790.190$, and potentially with the neighboring \ion{S}{v} $\lambda786.468$. As such, we were unable to find the column density of the resonance state \ion{O}{iv} 0, and could only find a lower limit of the \ion{O}{iv*} column density. We were also limited to finding a lower limit of the \ion{Ne}{viii} column density based on an AOD measurement due to the saturation of the doublet troughs. In the doublet of \ion{S}{xiv}, we determined the red trough of \ion{S}{xiv} $\lambda446$ to be contaminated, due to the visible blue-ward absorption compared to the blue trough $\lambda418$ (see Figure~\ref{fig:vcut2} plot c). Due to this limitation, we measured the AOD column density of \ion{S}{xiv} based on the blue trough. We determined that it was safe to treat this column density as a measurement, due to its shallower depth relative to similarly ionized troughs with comparable oscillator strengths (e.g. \ion{Si}{xii}).\par
\begin{figure}
    \centering
    \includegraphics[width=\linewidth]{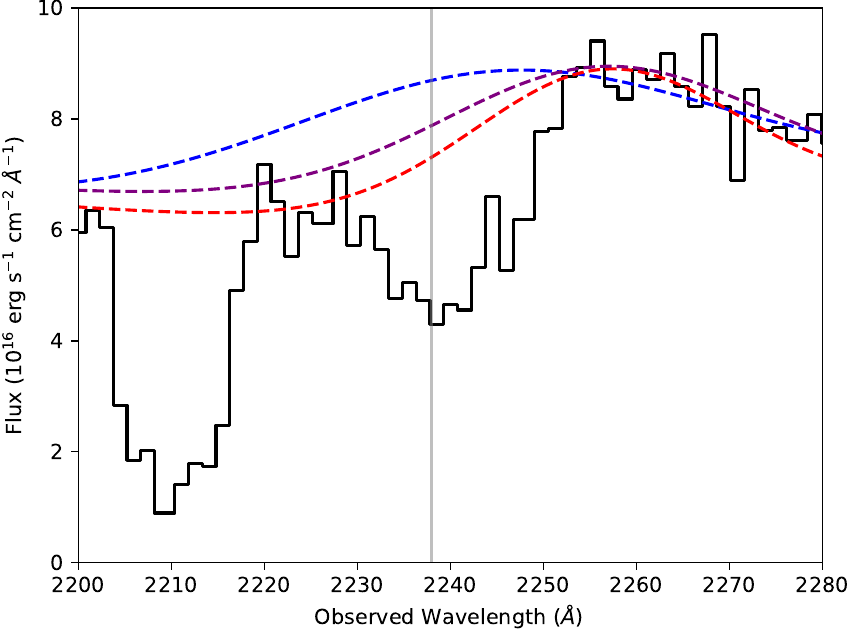}
    \caption{The \ion{O}{iv} region of the Q0254-334 spectrum showing the possible continuum fits which lead us to use $20\%$ error bars on the column densities (See text).}
    \label{fig:oiv_continuum}
\end{figure}

The integrated column densities are shown in Table~\ref{table:coldensity}. The rightmost column shows the values adopted for the photoionization solution described in Section~\ref{subsec:nvu}. The errors have been propagated from the error in the flux, and a conservative 20\% error has been added in quadrature to the adopted column density errors to account for the uncertainty in the continuum level due to the subjectivity of the model \citep{2018ApJ...858...39X}. This uncertainty is demonstrated in the column density calculation of \ion{O}{iv}* based on the different continuum models shown in Figure~\ref{fig:oiv_continuum}. The maximum, minimum, and intermediate continuum fits in the region are shown as blue, red, and purple dashed lines respectively. The \ion{O}{iv}* absorption is marked with a gray vertical line. The AOD measurement of the \ion{O}{iv}* column density is $28.0^{+2.8}_{-2.0}\times10^{14}$ cm$^{-2}$ for the intermediate continuum, while the higher and lower continuum levels yield results of $32.6^{+2.8}_{-2.0}\times10^{14}$ cm$^{-2}$ and $24.4^{+2.8}_{-2.0}\times10^{14}$ cm$^{-2}$. This indicates a $\pm15\%$ difference in column density depending on continuum level, and a $20\%$ difference when including the individual errors. Note that the column density of \ion{Ne}{v} was based on a Gaussian fit of the troughs of its different energy states, which we further describe in Section~\ref{subsec:nev}.\par
\begin{table}
	\centering
	\caption{Q0254-334 outflow column densities from STIS observations. The numbers next to the \ion{Ne}{v}* excited states denote the energies in cm$^{-1}$. The values are in units of $10^{14}$ cm$^{-2}$.}
	\label{table:coldensity}
	\renewcommand{\arraystretch}{1.4}
	\begin{tabular}{lcccc} % four columns, alignment for each
	\hline\hline
	Ion&AOD&PC&Adopted\\
	\hline
    \text{\ion{H}{i}} &$38_{-4}^{+7}$  &&$<38^{+11}$\\
    \text{\ion{C}{iii}} &$1.8_{-0.2}^{+0.3}$  &&$<1.8^{+0.4}$\\
    \text{\ion{N}{iv}} &$4.6_{-0.4}^{+0.6}$  &&$>4.6_{-1.0}$\\
    \text{\ion{O}{iv}*} &$28.0_{-2.0}^{+2.8}$  &&$>28.0{-6.0}$\\
    \text{\ion{O}{v}} &$50_{-5}^{+8}$  &&$>50_{-11}$\\
    \text{\ion{O}{vi}} &$114_{-4}^{+10}$  &&$>114_{-23}$\\
    \text{\ion{Ne}{v} total} &$196_{-22}^{+22}$  &&$196_{-45}^{+45}$\\
    \text{\ion{Ne}{v} 0} &$106_{-18}^{+18}$  &&\\
    \text{\ion{Ne}{v}* 411} &$72_{-12}^{+12}$  &&\\
    \text{\ion{Ne}{v}* 1109} &$18_{-3}^{+3}$  &&\\
    \text{\ion{Ne}{viii}} &$229_{-10}^{+25}$  &&$>229_{-47}$\\
    \text{\ion{Na}{viii}} &$29_{-2.9}^{+11.6}$  &&$>29_{-6.6}$\\
    \text{\ion{Na}{ix}} &$38.2_{-7.2}^{+8.4}$  &$48.2_{-6.1}^{+6.2}$&$48.2_{-11.3}^{+11.5}$\\
    \text{\ion{Mg}{x}} &$366_{-46}^{+87}$  &&$>336_{-87}$\\
    \text{\ion{Si}{xii}} &$360_{-22}^{+57}$  &&$>360_{-75}$\\
    \text{\ion{S}{iv} total} &$27^{+2}_{-2}$  &&$<27^{+5}$\\
    \text{\ion{S}{iv} 0} &$7.2^{+0.9}_{-0.6}$  &&\\
    \text{\ion{S}{iv}*} &$20^{+3}_{-2}$  &&\\
    \text{\ion{S}{xiv}} &$198_{-13}^{+20}$  &&$198_{-42}^{+45}$\\
    		\hline
	\end{tabular}
\end{table}
\subsection{\ion{Ne}{v} Gaussian Fitting}
\label{subsec:nev}
\begin{figure}
    \centering
    \includegraphics[width=\linewidth]{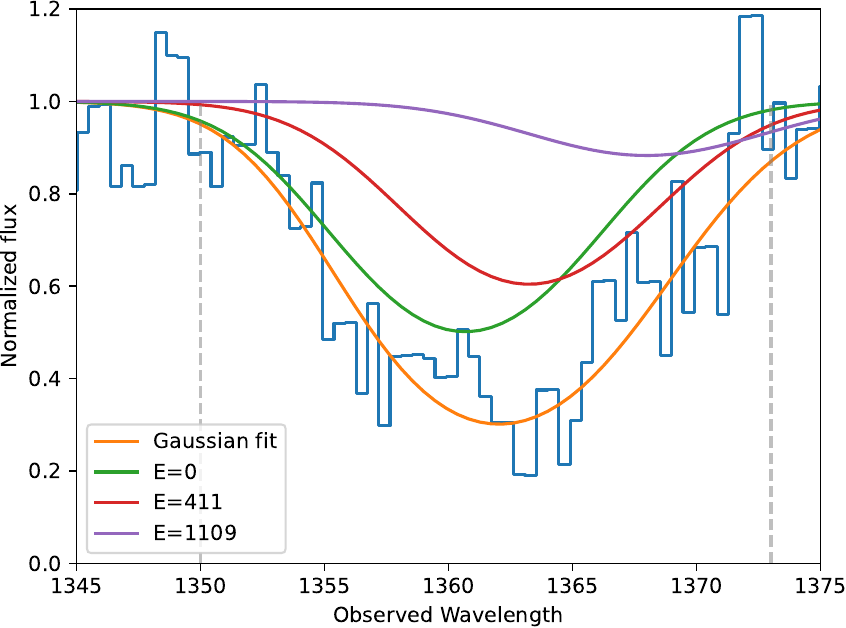}
    \includegraphics[width=\linewidth]{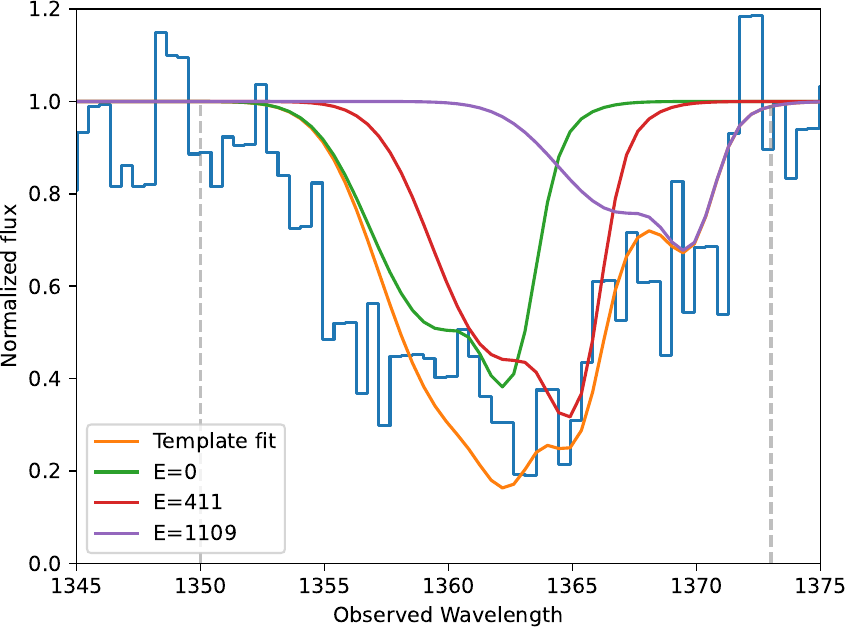}
    \caption{Modeling of the \ion{Ne}{v} absorption troughs, created by fitting Gaussians (top) and by using \ion{Si}{xii} as a template (bottom). The vertical dashed lines represent the range of data used for our fitting. The green curve is the modeled absorption of the resonance state of \ion{Ne}{v}. The red curve shows combined absorption of the $E=411\text{ cm${^-1}$}$ level lines, and the purple curve shows the absorption
    of the $E=1109\text{ cm$^{-1}$}$ level lines. The absorption features from multiple lines of the same excited states have been combined within the figure. The orange curve represents the total combined modeled absorption of the \ion{Ne}{v} multiplet. }
    \label{fig:nev_plot}
\end{figure}

As seen in Figure~\ref{fig:nev_plot} (top panel), the \ion{Ne}{v} multiplet of the outflow is blended into a singular trough. The involved transitions are \ion{Ne}{v} 0 ($\lambda=480.415$ \AA), \ion{Ne}{v}* 411 ($\lambda=481.227, 481.366, 481,371$ \AA), and \ion{Ne}{v}* 1109 ($\lambda=482.990, 482.994$ \AA). To remedy the blending, we modeled the individual energy states of \ion{Ne}{v} by fitting Gaussian profiles for each of the expected absorption features, and running a best fit algorithm to best match the data. The free parameters used were the optical depth of the ground state \ion{Ne}{v} trough, the width of the trough, and $\log{n_e}$. We assumed the AOD scenario, and adjusted the depths of the excited state troughs to match the oscillator strengths of the transition lines, as well as the abundance ratios $N(\ion{Ne}{v}*)/N(\ion{Ne}{v}\text{ }0)$ from the \textsc{Chianti} 9.0.1 atomic database \citep{1997A&AS..125..149D,Dere_2019}. We assumed a temperature of 10,000 K in our \textsc{Chianti} computations. A similar process of finding $n_e$ via the ratios between the different energy states of \ion{Ne}{v} is demonstrated by \citet{2020ApJS..247...39M}, and is especially illustrated in their Figure 3.\par
We have found that the optical depth $\tau=0.69\pm{0.11}$, $\text{FWHM}=2360\pm{170}\text{ km s}^{-1}$ and $\log{n_e}=3.6\pm{0.1} \text{ }[\text{cm }^{-3}]$. Using the modeled troughs, we have calculated the column densities of each energy state of \ion{Ne}{v}, as shown in Table~\ref{table:coldensity}. Since the value of $n_e$ is crucial in finding the distance of the outflow from the central source (as described in Section~\ref{sec:results}), we later ran a simulation with the spectral synthesis code \textsc{Cloudy} \citep[version c17.00,][]{2017RMxAA..53..385F} in order to verify the temperature of the outflow. With the two-phase high-ionization solution later described in Section~\ref{subsec:nvu} as our input parameters, the simulation yielded a temperature of $T\approx27,000\text{ K}$. Calculating the electron number density with this temperature yielded $\log{n_e}=4.0_{-0.1}^{+0.1}$. As such, we adopted this value of $\log{n_e}$ for the purpose of our analysis. The total column density of \ion{Ne}{v} based on this computation is in agreement with the value based on the $T=10,000$ K assumption.\par
\begin{figure}
    \centering
    \includegraphics[width=\linewidth]{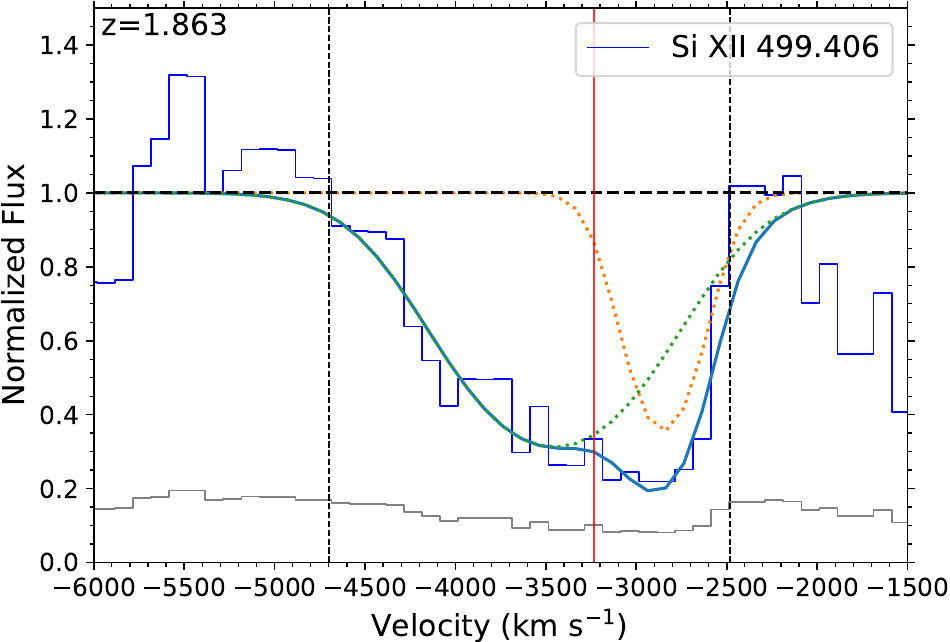}
    \caption{Fitting of a two-Gaussian profile to the absorption trough of \ion{Si}{xii}. The dotted curves show the individual Gaussians in the profile, while the blue curve shows the blended profile of both Gaussians. The dotted vertical lines represent the range of data that was used for fitting the Gaussians.}
    \label{fig:SiXII_fit}
\end{figure}

As an alternate method of modeling the blended trough of the \ion{Ne}{v} multiplet, we used the trough of \ion{Si}{xii} $\lambda499$ as a template to create a profile of two blended Gaussians (see Figure~\ref{fig:SiXII_fit}). We then ran a best fit algorithm to model the absorption of each energy state, leaving the width of the profile as a fixed parameter (see Figure~\ref{fig:nev_plot} bottom panel). This resulted in an electron number density of $\log{n_e}=4.3\pm0.1$ [cm$^{-3}$], which is only $\sim0.3$ dex higher than the simple Gaussian fitting shown in Section \ref{subsec:nev}. We report the physical properties calculated based on this value of $n_e$ in Table~\ref{table:energetics_siXII}.\par
While the difference in the electron number density shifts the kinetic luminosity to lower values relative to those shown in Table~\ref{table:energetics}, the kinetic luminosities remain in agreement within error. We thus focus on the results based on the Gaussian model throughout the paper. The parameters are described in further detail in Sections \ref{sec:results} and \ref{sec:discussion}.
\subsection{Photoionization Solution}
\label{subsec:nvu}
\begin{figure}
    \centering
    \includegraphics[width=\linewidth]{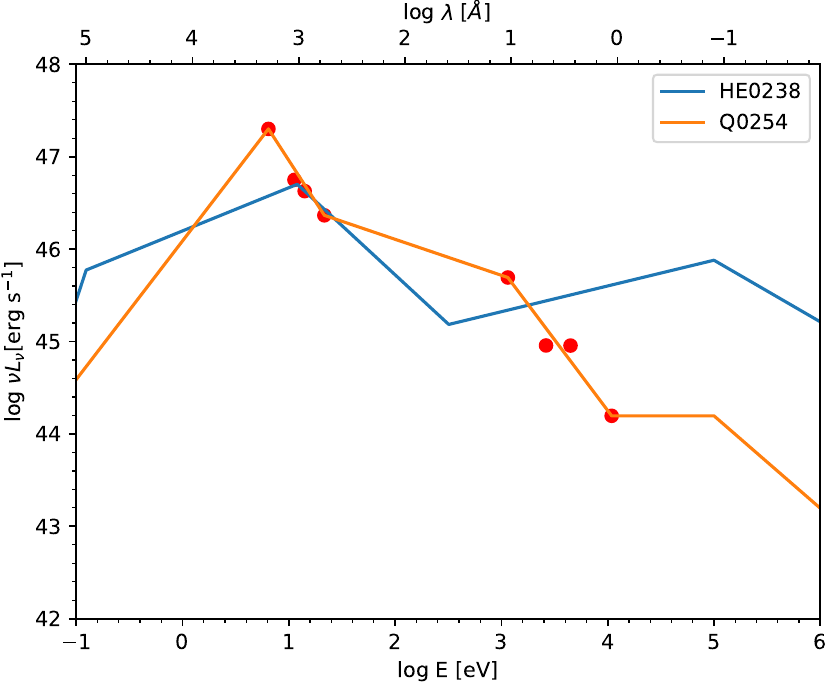}
    \caption{Comparison of SED shapes of HE0238 \citep{2013MNRAS.436.3286A} and Q0254-334. The Q0254-334 SED was formed by using the V-band magnitude \citep[the first red dot,][]{2006ApJ...636..610R}, UV continuum flux measured at three different wavelengths (rest wavelengths $\lambda=574, 880, 1097$ \AA; second, third, and fourth red dots), and the X-ray fluxes reported by Chandra (5th--8th dots). The HE0238 SED was scaled to match the UV continuum flux for the sake of this comparison. We use the Q0254-334 SED for the analysis in this paper.}
    \label{fig:sed_compare}
\end{figure}

\begin{figure}
    \centering
    \includegraphics[width=\linewidth]{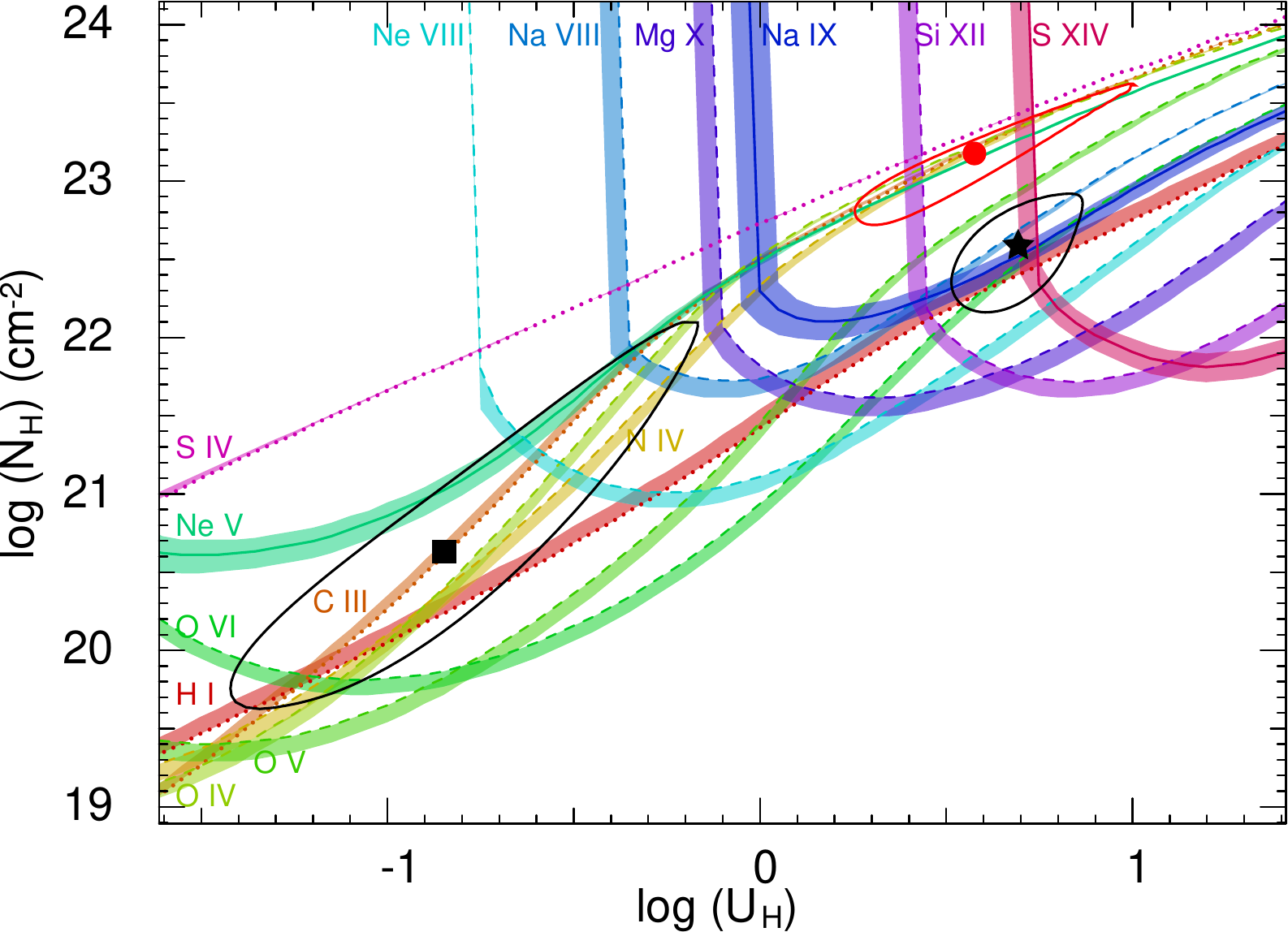}
    \includegraphics[width=\linewidth]{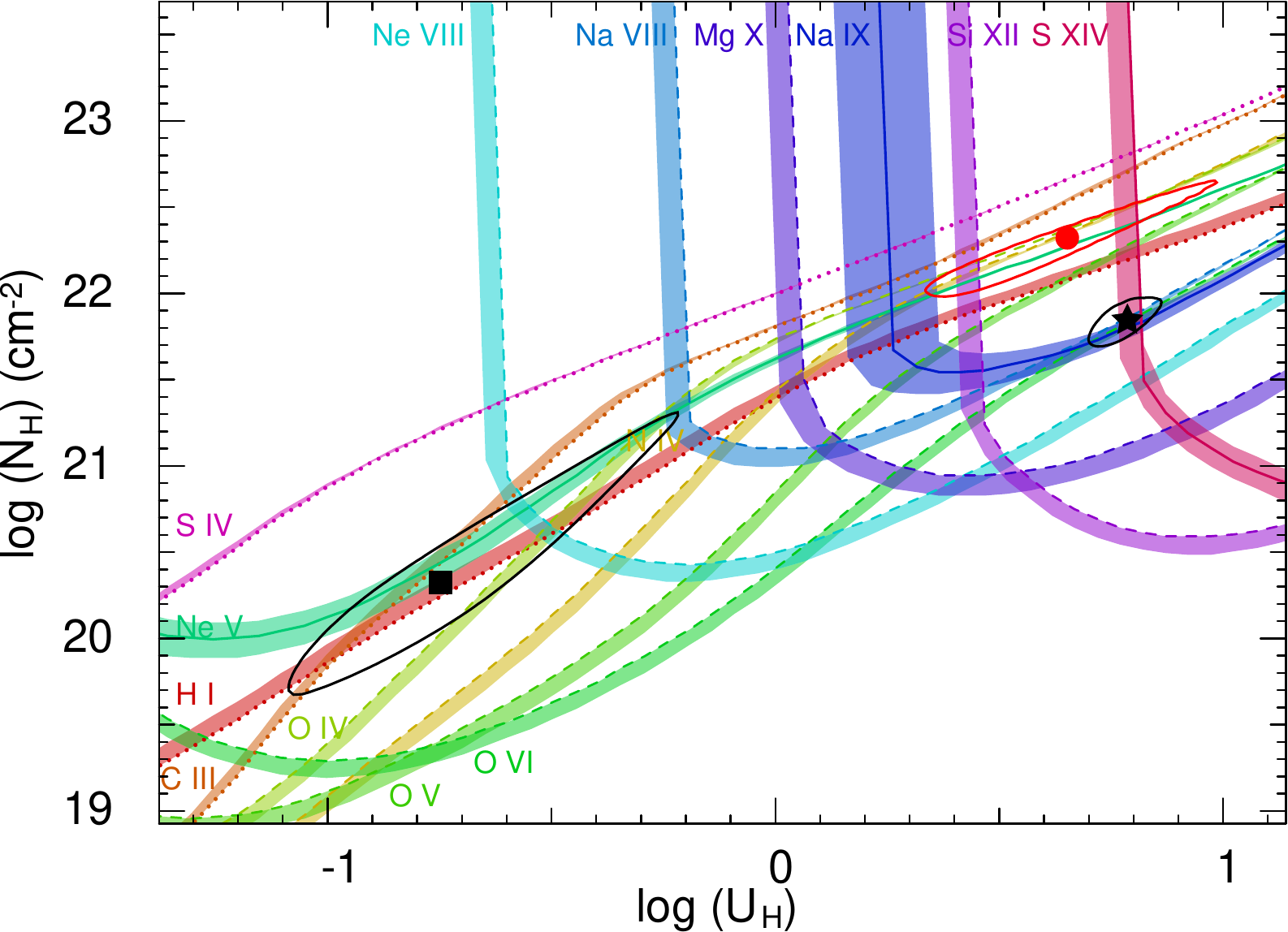}
    \caption{Plots of the Hydrogen column density ($N_H$) vs. ionization parameter ($U_H$), assuming solar (top) and supersolar (bottom) metallicities, with the Q0254-334 SED shown in Figure~\ref{fig:sed_compare}. Constraints on the parameters are based on measured column densities shown in Table~\ref{table:coldensity}. Measurements are shown as solid curves, while upper and lower limits are represented with dotted and dashed curves, respectively. The colored bands represent the uncertainties in the constraints. The red circle shows the one-phase solution of $N_H$ and $U_H$, while the black square and star show the high ionization and very high-ionization phase of the two-phase solution respectively. The 1-$\sigma$ uncertainties of the solutions are shown as black/red ellipses.}
    \label{fig:NVUplot}
\end{figure}

We used the measured ionic column densities to constrain the values of the hydrogen column density ($N_H$) and ionization parameter ($U_H$), as done in previous works \citep[e.g][]{2019ApJ...876..105X,10.1093/mnras/stac2194,10.1093/mnras/stac2638,2022Byun,2022MNRAS.516.3778W}. For this purpose, we used a grid of simulated models produced with \textsc{Cloudy} \citep{2017RMxAA..53..385F} with a range of $N_H$ and $U_H$ values as input parameters, modeling the ionic abundances at different $N_H$ and $U_H$. We used the ionic column densities shown in Table~\ref{table:coldensity} to set upper and lower limits to these parameters, as shown in Figure~\ref{fig:NVUplot}, assuming solar metallicity. We adopted a spectral energy distribution (SED) that would match the V-band flux of Q0254-334 found on NED, the UV continuum flux measured at three separate points, as well as the X-Ray fluxes observed with Chandra at energy ranges from 0.5--7 keV (see Figure~\ref{fig:sed_compare}). Note that there are limitations to this SED due to potential variability between the different observations that were referenced for its construction. \citet{2006ApJ...636..610R} report a V band magnitude of 16 and cite \citet{1982MNRAS.199...81W}, who in turn discuss observations made with the 3.9 m Anglo-Australian telescope on 28 November, 1978 and on 5 December 1978. Chandra observations of Q0254-334 were made on 2 January, 2000 and 15 February 2000. We have also calculated the $\alpha_{ox}$ spectral index based on our SED, using the following equation \citep{1979ApJ...234L...9T,Sobolewska_2009}:
\begin{equation}
    \alpha_{ox}=0.3838\log{\frac{L(2\text{ kev})}{L(2500\text{ \AA})}}
\end{equation}
which yielded a result of $\alpha_{ox}=-1.58$. This is somewhat higher than the range of $\alpha_{ox}$ values of LBQS broad absorption line quasars which were reported by \citet{2006ApJ...644..709G} (--2.58 to --1.65).\par
A single phase solution was insufficient to satisfy the constraints from the ionic column densities. To remedy this issue, we formulated a two-phase solution, in which a high- and very high-ionization phase exist co-spatially. We deduced that the two phases would be co-spatial based in the kinematic similarity between the high-ionization troughs and the very high-ionization troughs. Specifically, Figure~\ref{fig:nev_plot} shows that we get a very good fit for \ion{Ne}{v} (a high-ionization line), using the velocity template of \ion{Si}{xii} (a very high-ionization line, see Figure~\ref{fig:SiXII_fit}). We find that the two-phase solution satisfies more of the constraints set by the measured ionic column densities (reduced $\chi^2=5.1$, as opposed to $22.3$ for the one-phase solution). To cover the range of possible metallicities, we have also applied models of metallicity $Z\approx4.68Z_\odot$ \citep{2008A&A...478..335B,2020ApJS..247...41M}, which are shown in the lower panel of Figure~\ref{fig:NVUplot}. The results are favorable towards the super-solar metallicity solution, of which the reduced $\chi^2$ values are 16.0 and 0.5, for the one-phase and two-phase solutions respectively. As discussed by \citet{2013MNRAS.436.3286A}, the inability for a one-phase ionization solution to reasonably fit the measurements and limits of $N_{ion}$ necessitates the adoption of a two-phase solution. This is further demonstrated by comparing the modeled column densities of \ion{H}{i}, \ion{Na}{ix}, and \ion{S}{xiv} from supersolar one-phase and two-phase solutions to the observed ones, as shown in Table~\ref{table:ly_gamma}. Note that the reported measured column density of \ion{H}{i} is an upper limit based on \ion{Ly}{$\gamma$}. Due to the larger discrepancy between modeled and measured column densities in the solar abundance solutions (e.g. model \ion{H}{i} column density upwards of $\sim20$ times larger than measured), they have been excluded from the table. As can be seen in the table, the two-phase solution yields modeled column densities with a maximum $2\sigma$ difference between modeled and measured column densities, while the one-phase solution yields a $4-8\sigma$ difference. As such, the comparison of modeled column densities favors the two-phase solution. The $U_H$ and $N_H$ values found using $\chi^2$ analysis are shown in Table~\ref{table:energetics}.\par
We compared the $N_H$ and $U_H$ values found using the Q0254-334 SED with those found using the SED of the quasar HE0238-1904 \citep[hereafter HE0238,][]{2013MNRAS.436.3286A}, as the latter SED has been adopted for quasar outflow analysis in several past papers \citep[e.g.,][]{2020ApJS..247...39M,10.1093/mnras/stac2194,10.1093/mnras/stac2638,2022Byun,2022MNRAS.516.3778W}. We report the $\log{N_H}$ and $\log{U_H}$ values derived from the HE0238 SED in Table~\ref{table:he0238}. Comparing these values with those found in Table~\ref{table:energetics} shows that while the one-phase solutions are in agreement within error, the two-phase solutions show a discrepancy in the $\log{U_H}$ values that range up to $\sim0.5$ dex.
\begin{table}
	\centering
	\caption{The measured and modeled column densities of \ion{H}{i}, \ion{Na}{ix}, and \ion{S}{xiv} of the Q0254-334 outflow. The second and third columns denote the supersolar one-phase solution and the $\sigma$ difference between modeled and measured values; the fourth and fifth columns show the same for the two-phase solution. The values are in units of $10^{14}$ cm$^{-2}$.}
	\label{table:ly_gamma}
	\renewcommand{\arraystretch}{1.4}
	\begin{tabular}{lccccc} % four columns, alignment for each
	\hline\hline
	Ion&Measured&$4.68Z_\odot$1-p&$\Delta\sigma$ & $4.68Z_\odot$2-p&$\Delta\sigma$\\
	\hline
        \ion{H}{i}&$<38^{+11}$&123  &8&59&2\\
        \ion{Na}{ix}&$48^{+12}_{-11}$&96  &4&55&0.6\\
        \ion{S}{xiv}&$200^{+40}_{-40}$&40  &--4&190&--0.3\\
    		\hline
	\end{tabular}
\end{table}
\subsection{Black Hole Mass Calculation}
Black hole masses of AGN are often found using the emission features of \ion{Mg}{ii} \citep{2019ApJ...875...50B} or \ion{C}{iv} \citep{2006ApJ...641..689V,2017MNRAS.465.2120C}. However, as the STIS spectrum of Q0254-334 lacked both features, we looked to the \ion{O}{vi} emission to compute the mass of the central black hole. We referred to the method described by \citet{2013ApJ...774...67T}, measuring the \ion{O}{vi} FWHM to find the mass.\par
\begin{figure}
    \centering
    \includegraphics[width=\linewidth]{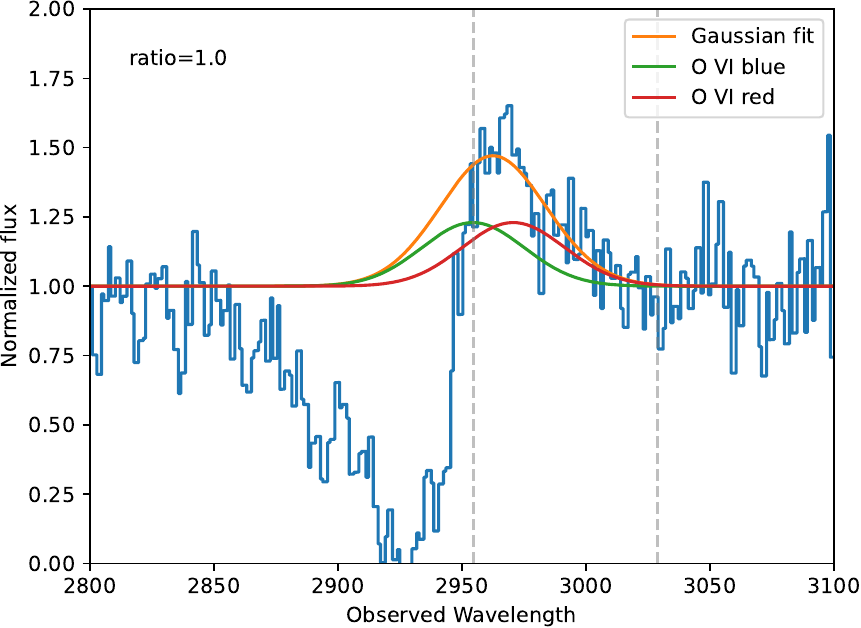}
    \caption{Gaussian fitting of the \ion{O}{vi} emission feature. The dashed vertical lines denote the range of data used for the Gaussian fit. The green and red curves are the modeled blue and red emission features respectively, and the orange curve represents the combined modeled emission.}
    \label{fig:ovi_plot}
\end{figure}

Although \citet{2013ApJ...774...67T} specify the use of two Gaussians to fit each line of the emission doublet, we opted to fit one Gaussian per line instead, as the lower signal to noise ratio of the STIS spectrum did not warrant the more detailed modeling method. We employed a best fit algorithm adjusting the amplitude of the blue emission line, the ratio between the blue and red lines, and the FWHM of the blue line. The ratio between the blue and red line amplitudes was constrained between 1--1.5, and the widths of the two features were fixed to be equal to each other. For the resulting fit, we found a ratio of 1, normalized amplitude $A=0.23\pm0.11$, and $\text{FWHM}=4800\pm900\text{ km s$^{-1}$}$. This, along with the measured flux of $F_{\lambda}=7.7^{+1.0}_{-1.0}\times10^{-16}\text{ erg s$^{-1}$ cm$^{-2}$ \AA$^{-1}$}$ at rest wavelength $1050$\AA, resulted in a black hole mass of $M_{BH}=5.3^{+5.5}_{-2.7}\times10^9 M_\odot$, and Eddington luminosity $L_{Edd}=6.6^{+6.9}_{-3.4}\times10^{47}\text{ erg s}^{-1}$. Note that we have limited our Gaussian fit to the red wing of the emission feature, as the blue wing has been contaminated by the absorption outflow (see Fig.~\ref{fig:ovi_plot}). While this contamination has contributed significantly to the uncertainty, we were unable to find alternative emission features with which to estimate the black hole mass.\par
\section{Discussion}
\label{sec:discussion}
\subsection{Distance of the Outflow from the Central Source}
\label{sec:results}
With the parameters we found as described in Section~\ref{sec:analysis}, we could calculate the distance of the outflow from the central source, as well as the kinetic luminosity of the outflow. The distance can be found based on the definition of the ionization parameter $U_H$:
\begin{equation}
    U_H\equiv \frac{Q_H}{4\uppi R^2 n_H c}
    \label{eq:uh}
\end{equation}
where $Q_H$ is the emission rate of ionizing photons, $R$ is the outflow distance from the source, $n_H$ is the hydrogen number density, and $c$ is the speed of light. Solving the equation for $R$ gives us
\begin{equation}
    R=\sqrt{\frac{Q_H}{4\uppi U_H n_H c}}
    \label{eq:distance}
\end{equation}
For highly ionized plasma, $n_e\approx 1.2 n_H$ \citep{2006agna.book.....O}, and the values of $U_H$ and $n_e$ were found in Section~\ref{sec:analysis}.\par
We followed the method of other works \citep[e.g.][]{2020ApJS..247...39M,10.1093/mnras/stac2194,2022Byun} to find $Q_H$ and integrated over the SED mentioned in Subsection~\ref{subsec:nvu}, limiting our range to energies over 1 Ryd. This yielded the bolometric luminosity $L_{Bol}=2.40^{+0.24}_{-0.24}\times10^{47}\text{ erg s$^{-1}$}$ and $Q_H= 9.33^{+0.94}_{-0.94}\times 10^{56}\text{ s$^{-1}$}$. The distance estimates of the outflow calculated with this value are shown in Tables~\ref{table:energetics} and \ref{table:energetics_siXII}.\par
\begin{table*}
	\centering
	\caption{Physical Properties of the Q0254-334 Outflow. The high and very high ionization phases for the two-phase solution are assumed to be co-spatial.}
	\label{table:energetics}
        \begin{threeparttable}
	\begin{tabular}{lcccccc}
	\hline\hline
        Metallicity & \multicolumn{3}{c}{$Z_\odot$} & \multicolumn{3}{c}{$4.68Z_\odot$}\\
	Solution &\text{One-phase}&\multicolumn{2}{c}{Two-phase}&\text{One-phase}&\multicolumn{2}{c}{Two-phase}\\
Phase&&High&Very High&&High&Very High\\
	\hline
\vspace{-0.2cm}$log(N_{\text{H}})$&\\\vspace{-0.2cm}
&$23.18^{+0.45}_{-0.46}$&$20.63^{+1.47}_{-1.01}$&$22.59^{+0.33}_{-0.43}$&$22.32^{+0.33}_{-0.34}$&$20.32^{+0.99}_{-0.65}$&$21.84^{+0.13}_{-0.15}$\\
$[\text{cm}^{-2}]$&\\
\hline
\vspace{-0.2cm}$log(U_{\text{H}})$&\\\vspace{-0.2cm} &$0.58^{+0.43}_{-0.32}$&$-0.85^{+0.68}_{-0.57}$&$0.69^{+0.17}_{-0.18}$&$0.65^{+0.33}_{-0.32}$&$-0.75^{+0.53}_{-0.34}$&$0.79^{+0.08}_{-0.09}$\\
$[\text{dex}]$&\\
\hline
\vspace{-0.2cm}$log(n_{\text{e}})$\tnote{*}&\\\vspace{-0.2cm} &$4.0^{+0.1}_{-0.1}$&$4.0^{+0.1}_{-0.1}$&$2.5^{+0.7}_{-0.6}$&$4.0^{+0.2}_{-0.2}$&$4.0^{+0.1}_{-0.1}$&$2.5^{+0.5}_{-0.4}$\\
$[\text{cm}^{-3}]$&\\
\hline
\vspace{-0.2cm}$\text{Distance}$&\\\vspace{-0.2cm}&$100^{+50}_{-40}$&\multicolumn{2}{c}{$500^{+490}_{-280}$}&$90^{+40}_{-30}$&\multicolumn{2}{c}{$450^{+240}_{-210}$}\\
$[\text{pc}]$&\\
\hline
\vspace{-0.2cm}$\dot M$&\\\vspace{-0.2cm}&$1400^{+1000}_{-700}$&\multicolumn{2}{c}{$1800^{+800}_{-300}$}&$180^{+90}_{-60}$&\multicolumn{2}{c}{$290^{+50}_{-90}$}\\
$[M_{\odot} \text{yr}^{-1}]$&\\
\hline
\vspace{-0.2cm}$\dot M v$&\\\vspace{-0.2cm}&$28^{+21}_{-14}$&\multicolumn{2}{c}{$37^{+16}_{-6}$}&$4^{+2}_{-1}$&\multicolumn{2}{c}{$6^{+1}_{-2}$}\\
$[10^{36} \text{ ergs cm}^{-1}]$&\\
\hline
\vspace{-0.2cm}$log({\dot E}_k)$&\\\vspace{-0.2cm}&$45.65^{+0.30}_{-0.24}$&\multicolumn{2}{c}{$45.78^{+0.07}_{-0.16}$}&$44.76^{+0.20}_{-0.18}$&\multicolumn{2}{c}{$44.98^{+0.15}_{-0.07}$}\\
$[\text{erg s}^{-1}]$&\\
\hline
\vspace{-0.2cm}${\dot E}_k/L_{Edd}$&\\\vspace{-0.2cm}&$0.7^{+1.1}_{-0.4}$&\multicolumn{2}{c}{$0.9^{+0.9}_{-0.5}$}&$0.08^{+0.10}_{-0.05}$&\multicolumn{2}{c}{$0.14^{+0.16}_{-0.07}$}\\
$[\text{\%}]$&\\
\hline
\vspace{-0.2cm}${\dot E}_k/L_{Bol}$&\\\vspace{-0.2cm}&$3.5^{+3.6}_{-1.5}$&\multicolumn{2}{c}{$4.6^{+1.0}_{-1.4}$}&$0.4^{+0.3}_{-0.2}$&\multicolumn{2}{c}{$0.7^{+0.3}_{-0.1}$}\\
$[\text{\%}]$&\\
	\hline
 
	\end{tabular}
  \begin{tablenotes}
     \item[*]The $n_e$ of the very high-ionization phase is the $n_e$ of the high phase times the ratio of the high/very-high ionization parameters.
 \end{tablenotes}
        \end{threeparttable}
\end{table*}
\begin{table*}
	\centering
	\caption{Photoionization solution for the Q0254-334 outflow assuming the HE0238 SED.}
	\label{table:he0238}
	\begin{tabular}{lcccccc}
	\hline\hline
        Metallicity & \multicolumn{3}{c}{$Z_\odot$} & \multicolumn{3}{c}{$4.68Z_\odot$}\\
	Solution &\text{One-phase}&\multicolumn{2}{c}{Two-phase}&\text{One-phase}&\multicolumn{2}{c}{Two-phase}\\
Phase&&High&Very High&&High&Very High\\
	\hline
\vspace{-0.2cm}$log(N_{\text{H}})$&\\\vspace{-0.2cm}
&$23.08^{+0.38}_{-0.38}$&$20.83^{+0.85}_{-0.59}$&$22.58^{+0.16}_{-0.19}$&$22.29^{+0.36}_{-0.34}$&$20.61^{+0.98}_{-0.94}$&$21.84^{+0.48}_{-0.19}$\\
$[\text{cm}^{-2}]$&\\
\hline
\vspace{-0.2cm}$log(U_{\text{H}})$&\\\vspace{-0.2cm} &$0.88^{+0.30}_{-0.24}$&$-0.39^{+0.47}_{-0.40}$&$1.00^{+0.06}_{-0.08}$&$0.87^{+0.29}_{-0.25}$&$-0.24^{+0.54}_{-0.59}$&$1.01^{+0.07}_{-0.09}$\\
$[\text{dex}]$&\\
\hline
	\end{tabular}

\end{table*}

\subsection{Contribution of the Outflow to AGN Feedback}
For an outflow to contribute to AGN feedback, its kinetic luminosity must be at least $\sim0.5\%$ \citep{2010MNRAS.401....7H} or $\sim5\%$ \citep{2004ApJ...608...62S} of the quasar's Eddington luminosity. Assuming an incomplete spherical shell, the mass flow rate can be calculated as follows:
\begin{equation}
    \dot{M}\simeq 4\uppi \Omega R N_H\mu m_p v
    \label{eq:mdot}
\end{equation}
followed by the kinetic luminosity:
\begin{equation}
    \dot{E}_k\simeq\frac{1}{2}\dot{M}v^2
    \label{eq:edotk}
\end{equation}
where $\Omega$ is the global covering factor, $\mu=1.4$ is the mean atomic mass per proton, $v$ is outflow velocity, and $m_p$ is the mass of a proton \citep{2012ApJ...751..107B}. We assumed $\Omega=0.2$, as \ion{C}{iv} BALs are found in $\sim20\%$ of quasars \citep{2003AJ....125.1784H}. We use the $\Omega$ associated with \ion{C}{iv} BALs, since our high-ionization phase has troughs from ions of very similar ionization potential. For example, in our spectrum, we detect \ion{O}{iv} $\lambda787$. \ion{O}{iv} has an ionization potential of 77 eV, which is quite similar to the \ion{C}{iv} ionization potential of 64 eV. Assuming supersolar metallicity, this calculation yielded a kinetic luminosity of $\log{\dot{E}_k}=44.76^{+0.20}_{-0.18} [\text{erg s$^{-1}$}]$ for the one-phase solution, and $\log{\dot{E}_k}=44.98^{+0.15}_{-0.07} [\text{erg s$^{-1}$}]$ for the two-phase solution, leaving a $\sim0.2$ dex difference between the solutions.\par
The ratio between the kinetic luminosity and Eddington luminosity yields $\dot{E}_k/L_{Edd}=0.08^{+0.10}_{-0.05}\%$ for the one-phase solution, and $\dot{E}_k/L_{Edd}=0.14^{+0.16}_{-0.07}\%$ for the two-phase solution, which is below the $0.5\%$ threshold. For the sake of completeness, we have also found the ratio between $\dot{E}_k$ and the bolometric luminosity $L_{Bol}$, resulting in $\dot{E}_k/L_{Bol}=0.4^{+0.3}_{-0.2}\%$ and $0.7^{+0.3}_{-0.1}\%$(see Table~\ref{table:energetics}). Based on the ratio between $\dot{E}_k$ and $L_{Edd}$, the outflow would be unable to contribute to AGN feedback. It is important to note that the different assumed metallicity values have significant effects on the physical parameters of the outflow, such as a near order of magnitude difference in kinetic luminosity, leading to values that may be sufficient for AGN feedback contribution (see Table~\ref{table:energetics}).\par
\subsection{The Two-Phase Outflow}
\label{subsec:twophase}
As mentioned earlier in Section \ref{subsec:nvu}, the two-phase photoionization solution provides a better fit to the constraints from the measured ionic column densities. While the values of $\dot{E}_k$ for the one-phase and two-phase solutions agree with each other within error (see Table~\ref{table:energetics}), there are significant differences to be found in the other parameters, such as distance, $N_H$, and $U_H$.\par
Note that the difference in $N_H$ between the high- and very high-ionization phases is $\sim1.5$ orders of magnitude, as well as the difference in $U_H$. Assuming the two phases are co-spatial, the volume filling factor of the high-ionization phase is as follows \citep{2013MNRAS.436.3286A,2020ApJS..247...39M}:
\begin{equation}
    f_\mathrm{V}=\frac{U_{H,HP}}{U_{H,VHP}}\times\frac{N_{H,HP}}{N_{H,VHP}}
\end{equation}
resulting in $\log{f_\mathrm{V}}=-3.1^{+1.1}_{-0.9}$, which follows our expectations from the high-ionization phase's larger $n_e$ and smaller $N_H$ values compared to those of the very high-ionization phase.\par
\subsection{Connection to X-Ray Warm Absorbers}
\label{subsec:xray}
The two-phase solution for the outflow of Q0254-334 is comparable to the parameters measured in X-ray warm absorbers. For instance, in their analysis of the Seyfert galaxy NGC 3783, \citet{2003ApJ...599..933N} found the parameters of the absorbing gas composed of three different components, with the oxygen ionization parameter ranging from $\log{U_{ox}}=-2.4$ to $-0.6$. To effectively compare the $U_H$ of Q0254-334 to the $U_{ox}$ values of NGC 3783, we calculated the oxygen ionizing emission rate $Q_{ox}$ as defined below:
\begin{equation}
    Q_{ox} = \int_{\nu (0.54 \text{ keV})}^{\nu (10 \text{ keV})}\frac{L_\nu}{h\nu}d\nu
\end{equation}
such that the ratio $\frac{Q_{ox}}{Q_H}=\frac{U_{ox}}{U_H}$. The resulting value of the emission rate was $Q_{ox}=3.9^{+0.4}_{-0.4}\times10^{54}\text{ s}^{-1}$, which is $2.4\pm{0.1}$ orders of magnitude smaller than $Q_H$. Subtracting $2.4\pm{0.1}$ from the $\log{U_H}$ values of the high- and very high-ionization phases leads to $\log{U_{ox}}= -3.2^{+0.5}_{-0.5}$ and $-1.6^{+0.3}_{-0.3}$ respectively. The very high-ionization phase has a $U_{ox}$ within the range of $U_{ox}$ values of the NGC 3783 absorbing gas. We note that NGC 3783 is a much lower luminosity AGN than Q0254-334, and that its SED may be different. However, lacking high quality X-ray spectra of $z\sim1$ quasars, it is still illuminating to compare the NGC 3783 X-ray wind with the EUV wind seen in Q0254-334.

\begin{table*}
	\centering
	\caption{Physical Properties of the Q0254-334 Outflow, based on the \ion{Si}{xii} template-based fitting of the \ion{Ne}{v} absorption. The high and very high ionization phases for the two-phase solution are assumed to be co-spatial.}
	\label{table:energetics_siXII}
	\begin{tabular}{lcccccc}
	\hline\hline
        Metallicity & \multicolumn{3}{c}{$Z_\odot$} & \multicolumn{3}{c}{$4.68Z_\odot$}\\
	Solution &\text{One-phase}&\multicolumn{2}{c}{Two-phase}&\text{One-phase}&\multicolumn{2}{c}{Two-phase}\\
Phase&&High&Very High&&High&Very High\\
	\hline
\vspace{-0.2cm}$log(n_{\text{e}})$&\\\vspace{-0.2cm} &$4.3^{+0.1}_{-0.1}$&$4.3^{+0.1}_{-0.1}$&$2.8^{+0.7}_{-0.6}$&$4.3^{+0.1}_{-0.1}$&$4.3^{+0.1}_{-0.1}$&$2.8^{+0.5}_{-0.3}$\\
$[\text{cm}^{-3}]$&\\
\hline
\vspace{-0.2cm}$\text{Distance}$&\\\vspace{-0.2cm}&$70^{+30}_{-20}$&\multicolumn{2}{c}{$370^{+350}_{-200}$}&$70^{+30}_{-20}$&\multicolumn{2}{c}{$330^{+170}_{-150}$}\\
$[\text{pc}]$&\\
\hline
\vspace{-0.2cm}$\dot M$&\\\vspace{-0.2cm}&$1000^{+700}_{-500}$&\multicolumn{2}{c}{$1300^{+500}_{-200}$}&$130^{+60}_{-50}$&\multicolumn{2}{c}{$210^{+30}_{-60}$}\\
$[M_{\odot} \text{yr}^{-1}]$&\\
\hline
\vspace{-0.2cm}$\dot M v$&\\\vspace{-0.2cm}&$20^{+15}_{-10}$&\multicolumn{2}{c}{$27^{+10}_{-3}$}&$2.6^{+1.3}_{-0.9}$&\multicolumn{2}{c}{$4.4^{+0.6}_{-1.2}$}\\
$[10^{36} \text{ ergs cm}^{-1}]$&\\
\hline
\vspace{-0.2cm}$log({\dot E}_k)$&\\\vspace{-0.2cm}&$45.52^{+0.30}_{-0.24}$&\multicolumn{2}{c}{$45.64^{+0.06}_{-0.15}$}&$44.63^{+0.19}_{-0.17}$&\multicolumn{2}{c}{$44.85^{+0.14}_{-0.06}$}\\
$[\text{erg s}^{-1}]$&\\
\hline
\vspace{-0.2cm}${\dot E}_k/L_{Edd}$&\\\vspace{-0.2cm}&$0.5^{+0.8}_{-0.3}$&\multicolumn{2}{c}{$0.6^{+0.6}_{-0.3}$}&$0.06^{+0.08}_{-0.03}$&\multicolumn{2}{c}{$0.10^{+0.11}_{-0.05}$}\\
$[\text{\%}]$&\\
\hline
\vspace{-0.2cm}${\dot E}_k/L_{Bol}$&\\\vspace{-0.2cm}&$2.6^{+2.6}_{-1.1}$&\multicolumn{2}{c}{$3.4^{+0.6}_{-1.0}$}&$0.33^{+0.19}_{-0.11}$&\multicolumn{2}{c}{$0.55^{+0.23}_{-0.08}$}\\
$[\text{\%}]$&\\
	\hline
	\end{tabular}
\end{table*}
\subsection{Comparison to Other Extreme UV Objects}
\label{subsec:euv500}
As the spectrum of Q0254-334 covers observed wavelengths as short as 400\AA, we found it appropriate to compare it with other quasars observed in the extreme UV range \citep[hereafter EUV500,][]{2020ApJS..247...37A}. We compiled a list of the physical parameters of 28 EUV500 quasar outflow systems analyzed in previous works \citep{2020ApJS..247...37A,2020ApJS..247...38X,2020ApJS..247...40X,2020ApJS..247...42X,2020ApJS..247...39M,2020ApJS..247...41M,2020ApJS..249...15M}, and added the parameters of Q0254-334 for comparison, with a total of 29 EUV500 outflow systems. Out of the 29 outflow systems, 24, including the outflow discussed in this paper, have measurements of kinetic luminosity and distance from the source. We compared the parameters of the Q0254-334 outflow such as $\dot{E}_K$, $N_H$, $R$, and $U_H$, with the other 23 outflow systems.\par
As seen in Figure~\ref{fig:euv_energetics}, no strong correlation has been found between $\log{\dot{E}_k}$ and $\log{L_{Edd}}$, or between $\log{\dot{E}_k}$ and $\log{L_{Bol}}$. 4 of the 24 outflows ($\sim16\%$) are above the threshold of $\dot{E}_k/L_{Edd}\sim5\%$, while 7 ($\sim29\%$) are between the $0.5\%$ and $5\%$ thresholds. With regards to $L_{Bol}$, 5 of the outflow systems ($\sim20\%$) are above the $5\%$ threshold, while 7 ($\sim29\%$) are between the $0.5\%$ and $5\%$ thresholds. Note that while the values of $\log{\dot{E}_k}$ range between 41--47, $\log{L_{Edd}}$ and $\log{L_{Bol}}$ range between 47.0--47.9 and 46.6--47.6 respectively, which is much narrower than the range of $\log{\dot{E}_k}$. It is also indicative of the ability of line-of-sight analysis to identify outflow systems at large ranges of kinetic luminosity, as well as velocity.
Figure~\ref{fig:euv_nvu} shows the $\log{N_H}$ and $\log{U_H}$ values of the high- and very high-ionization phases of each of the outflow systems. With the exception of the very high-ionization phase of the outflow system of UM425 traveling at $-9420\text{ km s$^{-1}$}$, the high-ionization phases tend to have values of $\log{U_H}<0$, while the very high-ionization phases have $\log{U_H}>0$. Note that the very high-ionization phase of the Q0254-334 outflow has a higher $\log{U_H}$ value relative to the average of the other outflows. We largely attribute this to the detection of \ion{S}{xiv}. As can be seen in Figure~\ref{fig:NVUplot}, the very high-ionization phase solution is at the intersection between the \ion{Na}{ix} and \ion{S}{xiv} constraints. The $\log{U_H}$ value is $\sim0.3$ dex higher than what it would have been if \ion{S}{xiv} were not detected, and the parameters were constrained by other ions such as \ion{Si}{xii}. It is also notable that as shown in Figure~\ref{fig:euv_nvu}, the $\log{U_H}$ of the very high-ionization phase of the Q0254-334 outflow is higher than the average $\log{U_H}$ of the other outflows. We suspect that future observed outflows with \ion{S}{xiv} would yield comparably high $U_H$ values.\par
\begin{figure}
    \centering
    \includegraphics[width=\linewidth]{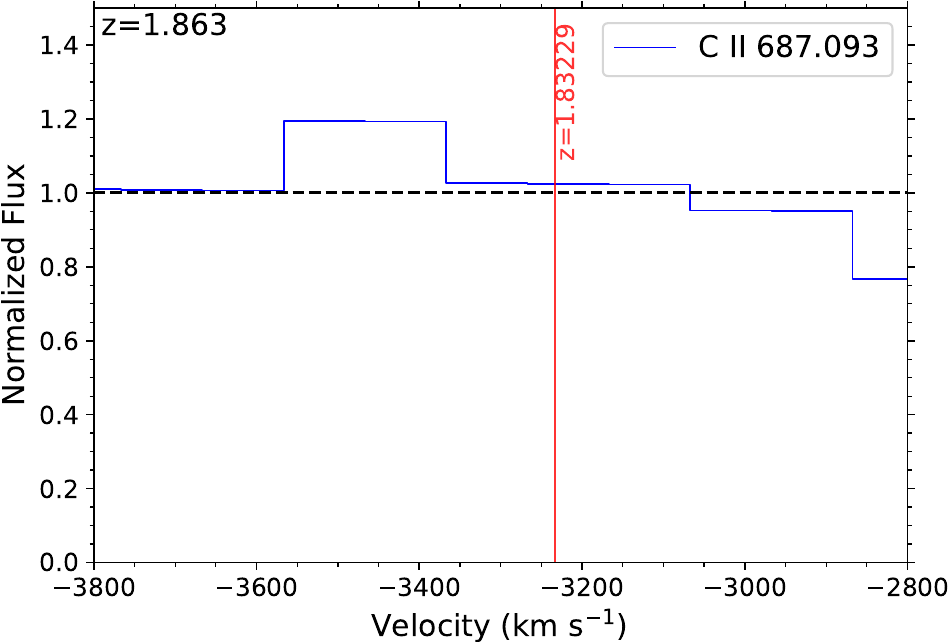}
    \caption{Region in which \ion{C}{ii} $\lambda687$ absorption is expected to be found.}
    \label{fig:CII}
\end{figure}
\begin{figure*}
    \centering
    \begin{multicols}{2}
    \includegraphics[width=\columnwidth]{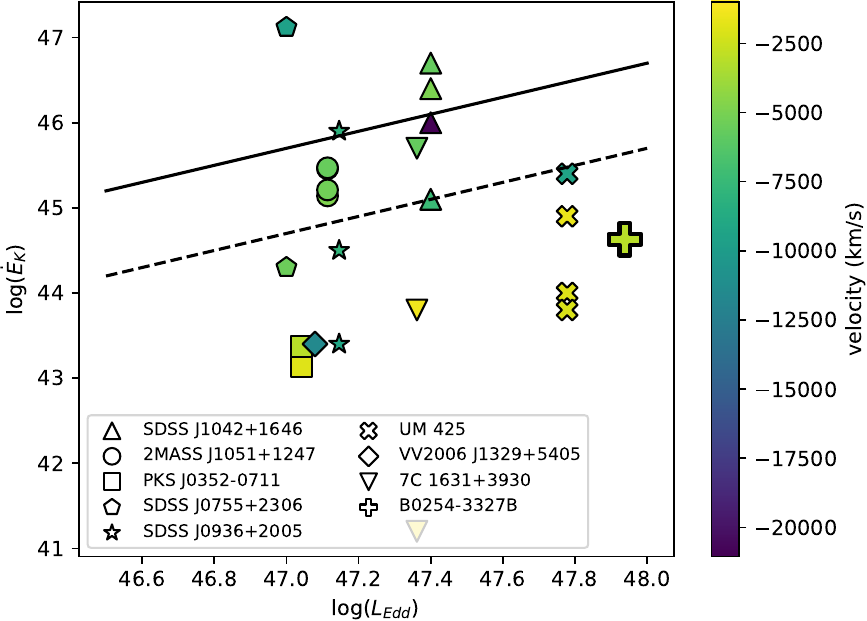}\par
    \includegraphics[width=\columnwidth]{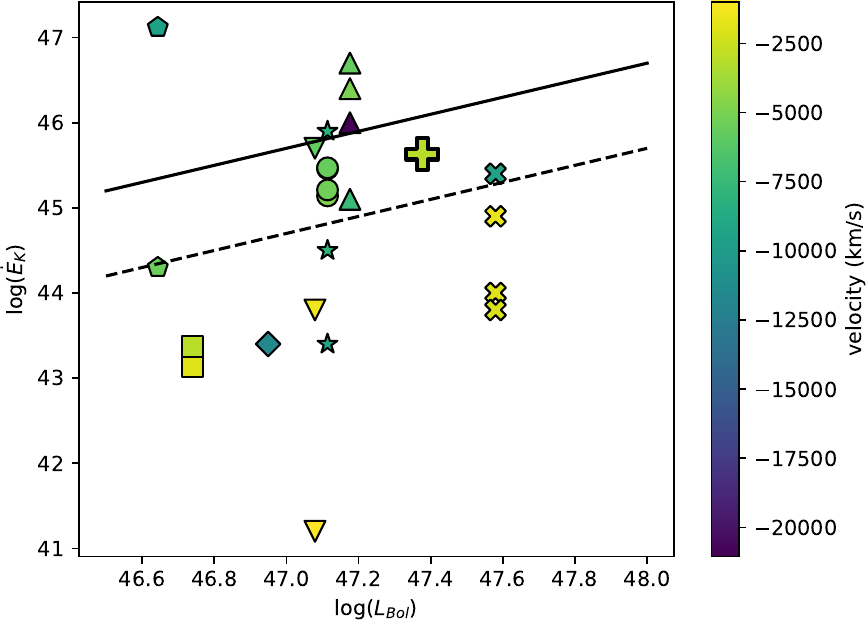}\par
    \end{multicols}
\caption{Distribution of $\log{\dot{E}_k}$ vs. $\log{L_{Edd}}$ (left) and $\log{L_{Bol}}$ (right) of EUV500 outflows. The dashed and solid lines on the left (right) indicate the $\dot{E}_k/L_{Edd}$ ($\dot{E}_k/L_{Bol}$) thresholds of $0.5\%$ and $5\%$ respectively. The plus-sign symbol denotes the outflow of Q0254-334 as reported in this paper, while the other symbols denote the parameters of other EUV outflows reported by \citet{2020ApJS..247...37A} and \citet{2020ApJS..249...15M}. The color map corresponds to the velocities of the outflow systems.}
\label{fig:euv_energetics}
\end{figure*}
\begin{figure}
    \centering
    \includegraphics[width=\columnwidth]{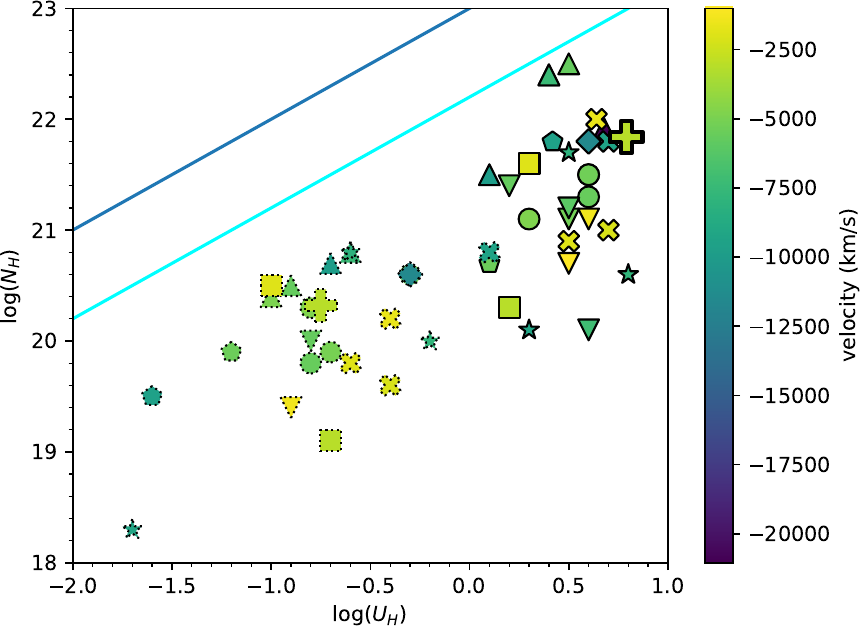}
    \caption{Distribution of $\log{N_H}$ vs. $\log{U_H}$ of EUV500 outflows. Symbols are coded as they are in Figure~\ref{fig:euv_energetics}. Symbols with dotted outlines denote high-ionization phases, while symbols with solid outlines denote very high-ionization phases. The color map corresponds to the velocities of the outflow systems. The dark blue and cyan curves show the \ion{H}{i} and \ion{He}{ii} ionization fronts respectively.}
    \label{fig:euv_nvu}
\end{figure}

Note that there is an apparent edge in the range of $\log{N_H}$ and $\log{U_H}$ values of the outflows. In Figure~\ref{fig:euv_nvu}, we have indicated the approximate locations of the hydrogen ionization front ($N_H\approx10^{23}U_H$), as well as the \ion{He}{ii} ionization front ($N_H\approx10^{22.2}U_H$). The $N_H/U_H$ ratio for the \ion{He}{ii} ionization front was calculated based on the average $\log{N_H}$ value at which the \ion{He}{ii} to \ion{He}{iii} ratio is 1:1 in a series of \textsc{Cloudy} models created with a range of $-2.0 < \log{U_H} < 1.0$. We used the aforementioned SED of HE0238 for the models, as this SED was used for the analysis of the majority of the EUV500 outflows in question. Interestingly, the $\log{N_H}$ vs. $\log{U_H}$ values of all of the EUV500 outflows fall under the \ion{He}{ii} ionization front, which would suggest that they are high ionized BALs (HiBALs) as opposed to low ionized BALs (LoBALs). This is supported by the lack of BALs from low-ionization species. For instance, there is a noticeable lack of absorption where the trough of \ion{C}{ii} $\lambda687$ would be, despite a large oscillator strength of f=0.336 (see Figure~\ref{fig:CII}).\par
\begin{figure}
    \centering
    \includegraphics[width=\columnwidth]{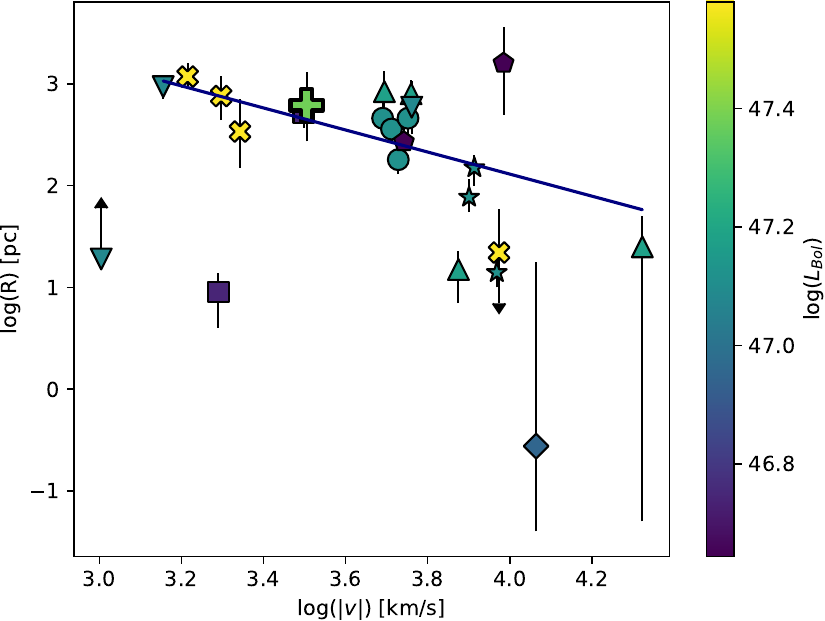}
    \caption{Distribution of $\log{R}$ vs. $\log{|v|}$ of EUV500 outflows. Symbols are coded as they are in Figure~\ref{fig:euv_energetics}, with errors in $\log{R}$ indicated with error bars. The dark blue line shows the weighted least squares linear fit of $\log{R}=-1.08\times\log{|v|}+6.44$, with adjusted $r^2=0.28$. The color map corresponds to the $\log{L_{Bol}}$ values of the outflows.}
    \label{fig:euv_distance}
\end{figure}

We also examined the ranges of $R$ and $v$ of the outflow systems (see Figure~\ref{fig:euv_distance}). To examine the correlation between distance and velocity, we conducted a weighted least squares linear fit between $\log{R}$ and $\log{|v|}$, taking into account the asymmetry of the reported errors in $R$. We adopted the weight determination method described by \citet{2003sppp.conf..250B}. The weight of each data point $w_i=1/V_i$ was determined by the value of $V_i$:
\begin{equation}
    V_i=\sigma_i^2+(1-\frac{2}{\uppi})\alpha_i^2
    \label{eq:V}
\end{equation}
in which $\sigma_i=\frac{\sigma_i^{+}+\sigma_i^{-}}{2}$ is the mean of the upper and lower errors of $\log{R}$, while $\alpha_i=\frac{\sigma_i^{+}-\sigma_i^{-}}{2}$.\par
The weighted linear fit yielded a slope of $-1.08$ and an intercept of $6.44$, suggesting a negative correlation. To determine the strength of the correlation, we calculated a modified value of the coefficient of determination $r^2$ that would take into account the weight of each data point. The residual sum of squares $SS_{res}$ was modified so that:
\begin{equation}
    SS_{res}= \sum_{i}w_i (\log{R}_i-f_i)^2
    \label{eq:SS_res}
\end{equation}
where $f_i$ is the value of $\log{R}$ according to the linear fit. The total sum of squares $SS_{tot}$ was adjusted so that:
\begin{equation}
    SS_{tot}= \sum_{i}w_i(\log{R}_i-\langle\log{R}\rangle)^2
    \label{eq:SS_tot}
\end{equation}
where $\langle\log{R}\rangle$ is the weighted mean of $\log{R}$. The resulting value of $r^2=1-\frac{SS_{res}}{SS_{tot}}$ is 0.28, suggesting a weak negative correlation between $\log{R}$ and $\log{|v|}$. This correlation is further supported by a Spearman correlation of -0.43 and associated p value of 0.05. We find it worth mentioning that \citet{2022ApJ...936..110C} have conducted a similar analysis of iron low ionized BALs (FeLoBALs), in which they determined that the correlation (or anti-correlation) between distance and velocity was dependent on the E1 parameter, which is formulated based on the relationship between the equivalent width of [\ion{O}{iii}] emission and the ratio between \ion{Fe}{ii} and \ion{H}{$\beta$} fluxes \citep{2022ApJ...935...92L}.
\section{Summary and Conclusion}
\label{sec:conclusion}
We have identified a BAL outflow in the HST/STIS spectrum of the quasar QSO B0254-3327B, of which we have found the ionic column densities (see Table~\ref{table:coldensity}). Based on the column densities, we conducted photoionization analysis to find the values of hydrogen column density $N_H$ and ionization parameter $U_H$. The results of our analysis are as follows:\par
\begin{enumerate}
    \item \textit{The two-phase solution.} The constraints from the measured ionic column densities required a solution with two ionization phases: the high-ionization phase and the very high-ionization phase. The two-phase solution showed a significant improvement in the $\chi^2$ value compared to that of the one-phase solution (reduced $\chi^2=5.1\text{ vs. }22.3$).
    \item \textit{The energetics of the outflow.} We were able to determine the composition of the \ion{Ne}{v} trough via Gaussian fitting of the blended features (see Figure~\ref{fig:nev_plot}), thanks to which we were able to narrow down the electron number density $n_e$. Through the use of Equations \ref{eq:distance}, \ref{eq:mdot}, and \ref{eq:edotk}, we were able to determine the distance, mass flow rate, and kinetic luminosity of the outflow (see Table~\ref{table:energetics}). There were notable differences in the energetics parameters based on different values of assumed metallicity ($Z=Z_\odot$ vs $Z=4.68Z_\odot$, see Tables~\ref{table:energetics} and \ref{table:energetics_siXII}).
    \item \textit{Potential contribution to AGN feedback.} As the ratio between the kinetic luminosity and the quasar's Eddington luminosity $\dot{E}_k/L_{Edd}=0.9^{+0.9}_{-0.5}\%$ assuming solar abundance, and $0.14^{+0.16}_{-0.07}$ assuming super-solar abundance, its contribution to AGN feedback is model dependent, as the theoretical thresholds for the ratio are $\sim0.5\%$ \citep{2010MNRAS.401....7H} and $\sim5\%$ \citep{2004ApJ...608...62S}.
    \item \textit{Comparison to X-ray warm absorbers.} We have compared the ionization parameter values of the high-ionization and very high-ionization phase to that of the X-ray warm absorber of NGC 3783 analyzed by \citet{2003ApJ...599..933N}. Converting $U_H$ to the oxygen ionization parameter $U_{ox}$ as defined by \citet{2003ApJ...599..933N} showed that the $U_{ox}$ of the very high-ionization phase agreed with that of the NGC 3783 absorber within the error.
    \item \textit{Comparison with other EUV500 outflows.}We have also examined the physical parameters of previously studied EUV500 outflows \citep{2020ApJS..247...37A,2020ApJS..247...38X,2020ApJS..247...40X,2020ApJS..247...42X,2020ApJS..247...39M,2020ApJS..247...41M,2020ApJS..249...15M}, and determined that out of the sample of 24 outflow systems with measured kinetic luminosity, up to $\sim50\%$ may contribute to AGN feedback, depending on the theoretical model. The trend between $\log{R}$ and $\log{|v|}$ was also analyzed via a weighted least squares linear fit, showing a weak negative correlation, indicated by a Spearman rank of --0.43, and p value of 0.05. We have also found that the very high-ionization phase of the Q0254-334 outflow had one of the highest $U_H$ values of all UV absorption outflows to date.
\end{enumerate}
The process of finding $\log{n_e}$ was limited to the examination of a blended trough of \ion{Ne}{v}. Further observations and analyses of the quasar may reveal more excited state troughs, which could help improve the uncertainty in $\log{n_e}$. Studying additional EUV500 outflows will be essential in a more thorough statistical analysis of their parameters as well.

\section*{Acknowledgements}

We thank Dr. J. Michael Shull and Dr. Kirk Korista for their input and advice, and Dr. Paola Rodriguez Hidalgo for her insight. We also thank the anonymous reviewer for their valuable input. We acknowledge support from NSF grant AST 2106249, as well as NASA STScI grants AR-15786, AR-16600, and AR-16601. This research has made use of the NASA/IPAC Extragalactic Database (NED), which is funded by the National Aeronautics and Space Administration and operated by the California Institute of Technology.

%%%%%%%%%%%%%%%%%%%%%%%%%%%%%%%%%%%%%%%%%%%%%%%%%%
\section*{Data Availability}

The STIS data of Q0254-334 described in this paper may be obtained from the MAST archive at \url{https://dx.doi.org/10.17909/63xd-4s51}. The FOS data shown for comparison can be obtained at \url{https://dx.doi.org/10.17909/z7ta-9048}.



\bibliographystyle{mnras}
\bibliography{Q0254} % if your bibtex file is called example.bib

%\begin{thebibliography}{99}
%\end{thebibliography}

%%%%%%%%%%%%%%%%%%%%%%%%%%%%%%%%%%%%%%%%%%%%%%%%%%

%%%%%%%%%%%%%%%%% APPENDICES %%%%%%%%%%%%%%%%%%%%%

%%%%%%%%%%%%%%%%%%%%%%%%%%%%%%%%%%%%%%%%%%%%%%%%%%

% Don't change these lines
\bsp	% typesetting comment
\label{lastpage}
\end{document}